\documentclass[a4paper,11pt]{article}
\pdfoutput=1 

\usepackage{jheppub} 

\usepackage{amsmath, mathtools}
\usepackage{amsfonts}
\usepackage{amssymb}
\usepackage{accents}
\usepackage{color}
\usepackage{slashed}
\usepackage{comment}
\usepackage{hyperref}
\usepackage{booktabs}
\usepackage{braket}
\usepackage{cellspace}
\usepackage{tikz}
\usepackage{pgfplots}
\usepackage{xcolor}

\usepgfplotslibrary{external}

\numberwithin{equation}{section}

\newcommand{\AdS}[1]{\text{AdS}_{#1}}
\newcommand{\CFT}{\text{CFT}}
\renewcommand{\S}{\text{S}}
\newcommand{\T}{\text{T}}

\newcommand{\AdSST}{\text{AdS}_3\times\text{S}^3\times\text{T}^4}

\renewcommand{\L}{\text{L}}
\newcommand{\R}{\text{R}}
\newcommand{\su}{\mathfrak{su}}
\renewcommand{\u}{\mathfrak{u}}
\newcommand{\psu}{\mathfrak{psu}}
\newcommand{\so}{\mathfrak{so}}
\renewcommand{\sl}{\mathfrak{sl}}

\newcommand{\de}{\text{d}}
\newcommand{\fe}{\text{f}}
\newcommand{\bo}{\text{b}}

\newcommand{\Htot}{H_{\text{tot}}}
\newcommand{\Ptot}{P_{\text{tot}}}
\newcommand{\Mtot}{M_{\text{tot}}}

\newcommand{\Reff}{R_{\text{eff}}}
\newcommand{\Neff}{N_{\text{eff}}}
\newcommand{\Nsc}{\mathcal{N}}

\newcommand{\NN}{N_{\text{eff}}}
\newcommand{\NNR}{\tilde N_{\text{eff}}}

\newcommand{\Qgen}{\mathbf{Q}}
\newcommand{\Lgen}{\mathbf{L}}
\newcommand{\Jgen}{\mathbf{J}}
\newcommand{\Hgen}{\mathbf{H}}

\newcommand{\Smat}{\mathbf{S}}
\newcommand{\Umat}{\mathbf{U}}

\title{\boldmath Integrable spin chain for stringy Wess-Zumino-Witten models}

\author{A.~Dei,}
\author{A.~Sfondrini}

\affiliation{Institut f\"ur theoretische Physik, ETH Z\"urich\\ Wolfgang-Pauli-Stra{\ss}e 27, 8093 Z\"urich, Switzerland}

\emailAdd{adei@itp.phys.ethz.ch}
\emailAdd{sfondria@itp.phys.ethz.ch}

\abstract{Building on \href{https://arxiv.org/abs/1804.01998}{arXiv:1804.01998} we investigate the integrable structure of the Wess-Zumino-Witten (WZW) model describing closed strings on $\AdS{3}\times\S^3\times\T^4$.
Using the recently-proposed integrable S~matrix we show analytically that all wrapping corrections cancel and that the theory has a natural spin-chain interpretation.
We construct the integrable spin~chain and discuss its relation with the WZW description. Finally we compute the spin-chain spectrum in closed form and  show that it matches the WZW prediction on the~nose.
}

\begin{document}
\maketitle
\flushbottom


\section{Introduction}

The correspondence between gravity on $\AdS{3}$ and conformal field theory in two dimensions~($\CFT_2$) is a key example of holographic duality~\cite{'tHooft:1993dmi,Susskind:1994vu}. From the seminal work of Brown and Henneaux~\cite{Brown:1986nw} to the remarkable developments in string theory~\cite{Maldacena:1997re}, see \textit{e.g.}\ ref.~\cite{David:2002wn} for a review, the $\AdS{3}/\CFT_{2}$ correspondence stands out for the possibility of performing \textit{exact computations} by conformal field theory techniques. In string theory such techniques arise in two distinct ways. Firstly, the dual conformal field theory is two-dimensional and, at weak string tension, is described by an almost-free CFT---more specifically, by the symmetric-product orbifold of a free CFT. Secondly, there exist superstring backgrounds that are supported by Neveu-Schwarz-Neveu-Schwarz (NS-NS) fluxes \textit{only}. These can be described by a $\CFT_2$ on the worldsheet, without the usual complications due to Ramond-Ramond (R-R) fluxes~\cite{Giveon:1998ns}. Such a $\CFT_2$ involves an $\sl(2,\mathbb{R})$ Wess-Zumino-Witten model describing (the chiral part of) $\AdS{3}$, which can be studied in detail following Maldacena and Ooguri~\cite{Maldacena:2000hw}, see also refs.~\cite{Pakman:2003cu,Israel:2003ry,Raju:2007uj,Giribet:2007wp, Ferreira:2017pgt}.

A more recent development is that strings on maximally supersymmetric $\AdS{3}$ backgrounds are classically \textit{integrable}~\cite{Babichenko:2009dk,Sundin:2012gc,Cagnazzo:2012se}, and indeed this integrability seems to carry over to the quantum theory when this is constructed in light-cone gauge, see ref.~\cite{Sfondrini:2014via} for a review (see also refs.~\cite{Arutyunov:2009ga,Beisert:2010jr} for broader reviews of AdS/CFT integrability). More specifically, an exact worldsheet scattering matrix was constructed for strings on R-R backgrounds~\cite{Borsato:2012ud,Borsato:2013qpa,Borsato:2014exa} and the relevant dressing factors were proposed~\cite{Borsato:2013hoa,Borsato:2016kbm,Borsato:2016xns}.
Quite remarkably, even backgrounds supported by a mixture of R-R and NS-NS fluxes are classically integrable~\cite{Cagnazzo:2012se} and their S~matrix can similarly be determined~\cite{Hoare:2013pma, Hoare:2013ida, Lloyd:2014bsa}, though no proposal exists for the dressing factors yet.
Curiously, until recently no proposal for the worldsheet S~matrix at the pure NS-NS point existed---despite the fact that, at least in the RNS formalism, the resulting theory is substantially simpler than a generic mixed-flux one.
Technically this is due to the fact that the light-cone symmetry algebra contracts at the NS-NS point.%
\footnote{More specifically, the off-shell central extension (first discussed in refs.~\cite{Arutyunov:2006ak, Arutyunov:2006yd} in the context of worldsheet integrability, see also~\cite{Beisert:2005tm}) vanishes at the pure-NS-NS point~\cite{Lloyd:2014bsa}.
}

Recently, in ref.~\cite{Baggio:2018gct}, an exact integrable worldsheet S~matrix for strings on $\AdSST$ \textit{with pure-NS-NS fluxes} was proposed by a different approach, based on $T\bar{T}$ deformations~\cite{Zamolodchikov:2004ce,Dubovsky:2013ira, Smirnov:2016lqw,Cavaglia:2016oda}. This S~matrix, including its dressing factor, is much simpler than its mixed-flux~\cite{Lloyd:2014bsa} (or even pure-R-R~\cite{Borsato:2014hja}) counterpart; in fact, it is very reminiscent of the worldsheet S-matrix of strings \textit{on flat space}~\cite{Dubovsky:2012wk}.
Ref.~\cite{Baggio:2018gct} and the simplicity of the S~matrix pave the way to a systematic investigation of strings on WZW backgrounds by integrability---establishing a firm contact between $\CFT_2$ and integrability techniques in $\AdS{3}/\CFT_2$.%
\footnote{%
Attempts have also been made to identify integrable structures in the \textit{dual} (rather than worldsheet) $\CFT_2$, see refs.~\cite{Pakman:2009mi,Sax:2014mea}.
}

The aim of this paper is to build on the proposal of ref.~\cite{Baggio:2018gct} and discuss the integrable structure of the WZW model for closed strings on~$\AdSST$ in greater detail. We claim that \textit{the WZW model can be described as a simple integrable spin~chain}. Even for theories that admit an integrable worldsheet S~matrix, this is a rather strong claim for at least two reasons:
\begin{enumerate}
\item An integrable spin-chain is a \textit{quantum-mechanical} system. The worldsheet theory of closed strings is (in an appropriate light-cone gauge) a two-dimensional \textit{quantum field theory}. Even in presence of integrability, the spin-chain description is usually only approximate, and is spoiled by the so-called wrapping corrections~\cite{Ambjorn:2005wa}.%
\footnote{In the context of $\AdS{3}/\CFT_2$ the effect of wrapping corrections is more severe than normally due to the presence of gapless excitations~\cite{Abbott:2015pps}.}
\item The WZW description gives a closed formula for the spectrum of strings on $\AdSST$. Generally the Bethe equations of an integrable spin-chain are hard to solve for generic states---even for relatively simple models, such as Heisenberg's chain. It is quite rare that a spin chain admits a closed-formula expression for its spectrum.
\end{enumerate}
We shall see that both of these issues can be addressed, building on the results presented in ref.~\cite{Baggio:2018gct}. Namely,
\begin{enumerate}
\item For this model,  wrapping corrections \textit{cancel exactly}. Hence, the number of particles is conserved microscopically, like for integrable spin~chains. To our knowledge, this is the only physical interacting finite-volume QFT$_{2}$ where this happens.%
\footnote{By extrapolating our results it is clear that the same must hold for superstrings in flat space, as well as on more general pure-NS-NS backgrounds. The physical reason for this is that this theory can be realised as a generalised $T\bar{T}$ deformation of a free theory~\cite{Baggio:2018gct}.}
\item The Bethe equations can indeed be solved exactly for arbitrary excited states, and match on the nose the WZW spectrum.
\end{enumerate}

Let us detail the plan of this paper.
We start by reviewing some properties of the light-cone gauge (Green-Schwarz) construction for strings on~$\AdSST$ supported by pure-NS-NS fluxes, and by summarising the claims of ref.~\cite{Baggio:2018gct} in section~\ref{sec:recap}. 
In section~\ref{sec:mTBA} we derive explicitly the cancellation of wrapping corrections within the mirror thermodynamic Bethe ansatz (mTBA) formalism. As section~\ref{sec:mTBA} is somewhat tangential to the construction of the spin~chain and to the study of its properties which we present immediately afterwards, some readers might want to overlook it in a first reading.
Next, in section~\ref{sec:spinchain} we construct the integrable spin~chain, and we show that the Bethe equations can be solved analytically, giving a closed formula for the spin~chain energy.
Finally in section~\ref{sec:WZWcomp} we explore how the spin~chain relates to the WZW construction. In particular, we find that the spin-chain magnons have a natural interpretation in terms of the modes of the WZW Ka\v{c}-Moody algebra---though this requires considering different ``spectrally flowed'' sectors, \textit{cf.}\ ref.~\cite{Maldacena:2000hw}. We conclude in section~\ref{sec:conclusions}.

In an effort to make this paper self-contained we present some review material in the appendices: the uniform light-cone gauge~\cite{Arutyunov:2004yx,Arutyunov:2005hd,Arutyunov:2006gs} for $\AdS{n}\times\S^n$  strings (appendix~\ref{app:lcgauge}), a technical point on the worldsheet S~matrix ``frames''~\cite{Arutyunov:2006yd,Sfondrini:2014via} (appendix~\ref{app:Smatframe}), the derivation of the mTBA equations equations for a non-relativistic theory of bosons and fermions with diagonal scattering (appendix~\ref{app:mTBA}) and some essential features of the WZW construction for strings on $\AdSST$ (appendix~\ref{app:WZW}). 

\section{Superstrings on \texorpdfstring{AdS$_{\bf3}\boldsymbol\times$S$^{\bf3}\boldsymbol\times$T$^{\bf 4}$}{AdS3xS3xT4}}
\label{sec:recap}
We start by briefly reviewing some features of strings on the pure-NS-NS $\AdSST$ background which motivate our construction, following refs.~\cite{Baggio:2018gct,Borsato:2014hja,Lloyd:2014bsa}.

\subsection{Symmetries of the background}
\label{sec:symmetries}
The isometries of the string non-linear sigma model are $\so(2,2)\oplus\so(4)_1\oplus\so(4)_2$, where $\so(4)_1$ corresponds to $\S^3$ isometries and has the interpretation of R-symmetry in the dual CFT, while $\so(4)_2$ corresponds to the four flat directions. This symmetry is spoiled by the boundary conditions of the $\T^4$ fields, but it is nonetheless useful to label fields and excitations. The background also possesses 16 Killing spinors. Eight of these close in the algebra $\psu(1,1|2)_{\L}$, where the label ``L'' stands for ``left'', whose anticommutation relations take the form
\begin{equation}
\begin{gathered}
\{\Qgen_{\pm}^{Aa},\,\Qgen_{\pm}^{Bb}\} =  \pm\epsilon^{AB}\epsilon^{ab} \Lgen^{\mp}\,,\qquad
\{\Qgen_{I}^{\pm a},\,\Qgen_{J}^{\pm b}\} =  \pm\epsilon_{IJ}\epsilon^{ab} \Jgen^{\pm}\,,\\
\{\Qgen_{\pm}^{Aa},\,\Qgen_{\mp}^{Bb}\} = \epsilon^{AB}\epsilon^{ab}\big(\Lgen^3\pm\Jgen^3\big)\,.
\end{gathered}
\end{equation}
Decomposing $\so(2,2)=\sl(2)_{\L}\oplus\sl(2)_{\R}$ and $\so(4)_1=\su(2)_{\L}\oplus\su(2)_{\R}$, we have that the generators $\{\Lgen^{I}\}_{I=\pm,3}$ give $\su(2)_{\L}$, while the generators $\{\Jgen^{A}\}_{A=\pm,3}$ give  $\su(2)_{\L}$.
There are eight more supercharges, which generate $\psu(1,1|2)_{\R}$; we denote the ``right'' generators with tildes, and the corresponding indices with dots, \textit{e.g.}\ $\tilde{\Qgen}^{\dot{A}a}_{\dot{I}}$. Notice that both left and right generators carry one index~$a$. This corresponds to the outer $\su(2)$ automorphism of $\psu(1,1|2)$. In $\AdSST$ this automorphism can be identified with a subalgebra of $\so(4)_2$. Specifically, we can decompose $\so(4)_2=\su(2)_\bullet\oplus\su(2)_\circ$ and identify the index~$a$ with the action of~$\su(2)_\bullet$~\cite{Borsato:2014hja}.

\begin{table}[t]
\centering
\renewcommand{\arraystretch}{1.2}
\begin{tabular}{|c|c|c|c|}
\hline
Chiral part. & $\mu$ & $\ell-j$ & $-j$ \\
\hline 
$Y(p)$ & $1$ & $\tfrac{k}{2\pi}p+1$ & $1$\\
$\eta^a(p)$ & $1$ & $\tfrac{k}{2\pi}p+1$& $ \tfrac{1}{2}$ \\
$Z(p)$ & $1$ & $\tfrac{k}{2\pi}p+1$ & 0  \\
\hline
$\bar Z(p)$ & $-1$ & $\tfrac{k}{2\pi}p-1$& 0 \\
$\bar \eta^a(p)$ & $-1$& $\tfrac{k}{2\pi}p-1$ & $- \tfrac{1}{2}$ \\
$\bar Y(p)$ & $-1$ & $\tfrac{k}{2\pi}p-1$ & $-1$   \\
\hline
$\chi^{\dot{a}}(p)$ & $0$& $\tfrac{k}{2\pi}p$ & $\tfrac{1}{2}$\\
$T^{a\dot{a}}(p)$ & $0$& $\tfrac{k}{2\pi}p$ & $0$\\
$\bar{\chi}^{\dot{a}}(p)$ & $0$& $\tfrac{k}{2\pi}p$ & $-\tfrac{1}{2}$\\
\hline
\end{tabular}
\hspace*{1cm}\begin{tabular}{|c|c|c|c|}
\hline
Anti-chiral part. & $\mu$ & $\tilde{\ell}-\tilde{\jmath}$ & $-\tilde{\jmath}$ \\
\hline 
$Y(p)$ & $1$ & $-\tfrac{k}{2\pi}p-1$ & $-1$\\
$\eta^a(p)$ & $1$ & $-\tfrac{k}{2\pi}p-1$& $ -\tfrac{1}{2}$ \\
$Z(p)$ & $1$ & $-\tfrac{k}{2\pi}p-1$ & 0  \\
\hline
$\bar Z(p)$ & $-1$ & $-\tfrac{k}{2\pi}p+1$& 0 \\
$\bar \eta^a(p)$ & $-1$& $-\tfrac{k}{2\pi}p+1$ & $ \tfrac{1}{2}$ \\
$\bar Y(p)$ & $-1$ & $-\tfrac{k}{2\pi}p+1$ & $1$   \\
\hline
$\chi^{\dot{a}}(p)$ & $0$& $-\tfrac{k}{2\pi}p$ & $-\tfrac{1}{2}$\\
$T^{a\dot{a}}(p)$ & $0$& $-\tfrac{k}{2\pi}p$ & $0$\\
$\bar{\chi}^{\dot{a}}(p)$ & $0$& $-\tfrac{k}{2\pi}p$ & $\tfrac{1}{2}$\\
\hline
\end{tabular}
\renewcommand{\arraystretch}{1}
\caption{%
The particle spectrum of pure-NS-NS $\AdSST$ Green-Schwarz strings. We have eight bosons: two correspond to transverse modes on $\AdS{3}$ ($Z,\bar{Z}$), two to transverse modes on $\S^3$ ($Y,\bar{Y}$), and four to the torus~($T^{a\dot{a}}$). The latter transform under $\so(4)_2=\su(2)_\bullet\oplus\su(2)_\circ$. We group them in blocks with the fermions, corresponding to four $({\bf 2}|{\bf 2})$ supersymmetric representations~\cite{Lloyd:2014bsa}; notice that all torus excitations carry an index $\dot{a}\sim\su(2)_\circ$. Particles in the same multiplet have the same dispersion~\eqref{eq:dispersion}, but potentially different R~charge~$j$ (or~$\tilde{\jmath}$). Since particles are chiral, we list separately the case in which $\partial_p H>0$ (left table, chiral particles) and the one where $\partial_p H<0$ (right table, anti-chiral particles). In the former case, the light-cone energy is entirely given by left-charges $\ell-j$ (while $\tilde{\ell}=\tilde{\jmath}=0$), and \textit{viceversa} in the latter. Notice that charges also ``flip sign'' when changing the momentum of a particle from the chiral region ($p>-2\pi\mu/k$) to the anti-chiral one ($p<-2\pi\mu/k$), \textit{i.e.}, $j\leftrightarrow-\tilde{\jmath}$. This is because $\mu$, which is the difference of left and right charges, remains constant.
}
\label{tab:excitations}
\end{table}
\subsection{Light-cone gauge symmetries and representations}
Integrability manifests itself when quantising the theory in light-cone gauge, much like in the $\AdS{5}\times\S^5$ case~\cite{Arutyunov:2006ak,Arutyunov:2006yd}. We fix light-cone gauge by picking a 1/2-BPS geodesic with R~charge $R=j+\tilde{\jmath}$ such that%
\footnote{%
Here and below we indicate with $j$ the eigenvalues of $\Jgen^3$ and with $\ell$ the eigenvalue of~$\Lgen^3$, and similarly for the right charges.
}
\begin{equation}
j=\tilde{\jmath}=\ell=\tilde{\ell}\,,
\end{equation}
see also appendix~\ref{app:lcgauge}.
This breaks $\so(2,2)\oplus\so(4)_1$ to its Cartan elements and preserves only half of the supercharges. The light-cone Hamiltonian is
\begin{equation}
\label{eq:lcHam}
\Hgen = \Lgen^3 - \Jgen^3 + \tilde{\Lgen}^3 - \tilde{\Jgen}^3\,,
\end{equation}
which is positive semi-definite owing to the $\psu(1,1|2)_\L\oplus \psu(1,1|2)_\R$ BPS bounds, $\Lgen^3 \geq \Jgen^3$ and $\tilde{\Lgen}^3 \geq \tilde{\Jgen}^3$. In a suitable light-cone gauge (see appendix~\ref{app:lcgauge}) the length of the worldsheet~$R$ is given by the R~charge. After gauge~fixing we are left with eight bosonic and eight fermionic degrees of freedom. As discussed at length in refs.~\cite{Borsato:2014hja,Lloyd:2014bsa}, these transform in \textit{short representations} of the residual supersymmetry algebra; the shortening condition can be expressed in terms of a quadratic constraint on $\Hgen$. This is very simple for pure-NS-NS backgrounds, and takes the form%
\footnote{The dispersion relation for backgrounds with NS-NS background fluxes can  also be found from studying giant-magnon solutions~\cite{Hoare:2013lja}.}
\begin{equation}
\label{eq:dispersion}
H(p,\mu) = \left|\frac{k}{2\pi}p+\mu\right|\,,
\end{equation}
where $k\in\mathbb{N}$ is the WZW level%
\footnote{%
The WZW level is proportional to the string tension; more precisely $k=R_{\text{AdS}}^2/\alpha'$.
}
 and $p$ is the worldsheet momentum. The shift $\mu$ is fixed by symmetry for different modes; we collect the 8 bosonic and 8 fermionic excitations in table~\ref{tab:excitations}. As discussed in refs.~\cite{Borsato:2014hja,Lloyd:2014bsa}, these fundamental excitations transform in four $({\bf 2}|{\bf 2})$ irreducible representation of the light-cone gauge symmetry algebra.
Notice the absolute value in eq.~\eqref{eq:dispersion} that signals that the theory is chiral on the worldsheet, even in light-cone gauge. For this reason we introduce the notion of \textit{left- and right-movers} on the worldsheet. These not to be confused with the ``L,R'' labels introduced above; to avoid such a confusion we will reserve the words ``chiral'' and ``anti-chiral'' to denote worldsheet left-/right-movers, and use ``left'' and ``right'' to denote the target-space charges. More specifically, we say that
\begin{equation}
\label{eq:chiralities}
\frac{\partial}{\partial p}H(p,\mu) =
\begin{cases}
\displaystyle
+\frac{k}{2\pi} &\text{chiral excitations,}\\[0.2cm]
\displaystyle
-\frac{k}{2\pi} &\text{anti-chiral excitations.}
\end{cases}
\end{equation}
The zero-modes of the energy requires a slightly more careful discussion which we will present later, see section~\ref{sec:spinchaincomparison}.

\subsection{S~matrix, Bethe-Yang equations and (no) finite-size corrections}
In the limit where the size of the worldsheet goes to infinity (\textit{i.e.}, for states of large R~charge), one can define a scattering matrix. For the $2\to2$ scattering of fundamental excitations, this gives a $16^2\times 16^2$ matrix. Building on the integrability of the underlying classical non-linear sigma model~\cite{Babichenko:2009dk, Sundin:2012gc,Cagnazzo:2012se}, it was shown in refs.~\cite{Borsato:2013qpa,Borsato:2014hja,Lloyd:2014bsa} that the $\AdS{3}\times\S^3\times\T^4$ S~matrix is consistent with scattering factorisation~\cite{Zamolodchikov:1978xm} and hence with integrability. Moreover, the $2\to2$ scattering matrix could be fixed (up to the so-called dressing factors~\cite{Borsato:2013hoa,Borsato:2016xns}) from the light-cone symmetries for pure-R-R and mixed-flux backgrounds. Unfortunately, this is not possible for pure-NS-NS backgrounds; while there is no reason to suspect a breakdown of quantum integrability at the NS-NS point,%
\footnote{If anything, we would suspect that this case is ``more symmetric'' than a generic mixed-flux background, as it can be described as a WZW model on the worldsheet.}
 symmetry arguments alone do not allow us to fix the two-particle S-matrix.
 
In ref.~\cite{Baggio:2018gct} it was proposed that the two-particle S~matrix of pure-NS-NS backgrounds is integrable and in fact~\emph{proportional to the identity}, \textit{i.e.}\ given entirely by a CDD factor. In particular, defining%
\footnote{%
CDD factors of this form were first considered in ref.~\cite{Arutyunov:2006gs} in the context of uniform light-cone ``$a$-gauge'' transformations, see also appendix~\ref{app:lcgauge}.
}
\begin{equation}
\label{eq:cddfactor}
\Phi(p_1,p_2)= \dfrac{1}{2} \left(p_1 H_2-p_2 H_1 - p_1m_2+p_2m_1 \right) \,,
\qquad
m_j = \mu_j\, \text{sgn}\big(\frac{k}{2\pi}p_j+\mu_j\big)\,,
\end{equation}
the \textit{exact S~matrix}%
\footnote{In the near-BMN limit, this proposal matches the tree-level result of ref.~\cite{Hoare:2013pma}.}
 is, in a suitable gauge and frame (see appendix~\ref{app:Smatframe}), 
\begin{equation}
\label{eq:Smatrix}
\Smat(p_1,p_2)=e^{i \Phi(p_1,_2)}\, \mathbf{1}\,.
\end{equation} 
The phase shift~\eqref{eq:cddfactor} can be written quite explicitly by taking into account the worldsheet chirality of the two particles, \textit{cf.}\ eq.~\eqref{eq:chiralities},
\begin{equation}
\Phi(p_i,p_j)=\begin{cases}
\displaystyle
\phantom{+}0 & p_i \text{ and } p_j \text{ both chiral or both anti-chiral},\\[0.1cm]
\displaystyle
-\frac{k}{2\pi} p_ip_j & p_i\text{ chiral and } p_j \text{ anti-chiral},\\[0.2cm]
\displaystyle
+\frac{k}{2\pi} p_ip_j & p_i\text{ anti-chiral and }p_j \text{ chiral}.
\end{cases}
\end{equation}
Notice that the phase-shift is completely independent of $\mu_i,\mu_j$ and coincides with the one occurring for strings in flat space~\cite{Dubovsky:2012wk}.

As the S~matrix is diagonal it is immediate to write down the Bethe-Yang equations, which  for a state with $K$ excitations are
\begin{equation}
\label{eq:BetheYang}
1=e^{i p_j\,R_0}\prod_{k\neq j}^K e^{i \Phi_{jk}(p_i,p_j)}=1\,,\qquad j=1,\dots K \, ,
\end{equation}
where $R_0$ is the charge of a reference (BPS) vacuum state, see also appendix~\ref{app:Smatframe}. These equations are supplemented by the level-matching constraint
\begin{equation}
\Ptot=\sum_i^K p_i = 2\pi W\,,
\end{equation}
which depends on the light-cone winding number~$W\in\mathbb{Z}$, see also eq.~\eqref{eq:windinglc}.

It is well understood~\cite{Ambjorn:2005wa} that the Bethe-Yang equations do not generally yield the correct spectrum of AdS/CFT integrability---in fact, the same is true for any finite-volume integrable system. It is necessary to properly account for finite-size ``wrapping corrections'' of the type described by L\"uscher~\cite{Luscher:1985dn,Luscher:1986pf}. This gives an infinite tower of corrections which can collectively be described within the (mirror) thermodynamic Bethe anstaz (TBA) formalism~\cite{Zamolodchikov:1989cf}. Following ref.~\cite{Baggio:2018gct}, we shall see that these wrapping corrections cancel for pure-NS-NS $\AdSST$ strings.
The reason for this striking result is twofold: on the one hand, the structure of scattering is extremely simple here; on the other, the fundamental excitations of table~\ref{tab:excitations} fall into supersymmetric multiplets.
Let consider a generic state with $K$ particles of arbitrary flavour $j_1, \dots j_K$ and momenta $p_1, \dots p_K$. Schematically, L\"uscher ``F-term'' is%
\footnote{%
Strictly speaking, we would need to use the ``string frame'' S~matrix to compute this integral, see appendix~\ref{app:Smatframe}. The argument would go through in exactly the same manner.
}
\begin{equation}
\label{eq:fterm}
\int \de u \frac{\partial \bar{p}(u)}{\partial u} e^{- \bar{H}(u)\, R}\sum_{X}(-1)^{F_X}
 S_{Xj_1}(\bar{p}(u),p_1)\, S_{Xj_2}(\bar{p}(u),p_2)\cdots
  S_{Xj_K}(\bar{p}(u),p_K)\,.
\end{equation}
\begin{figure}
  \centering
    \includegraphics[width=0.2\textwidth]{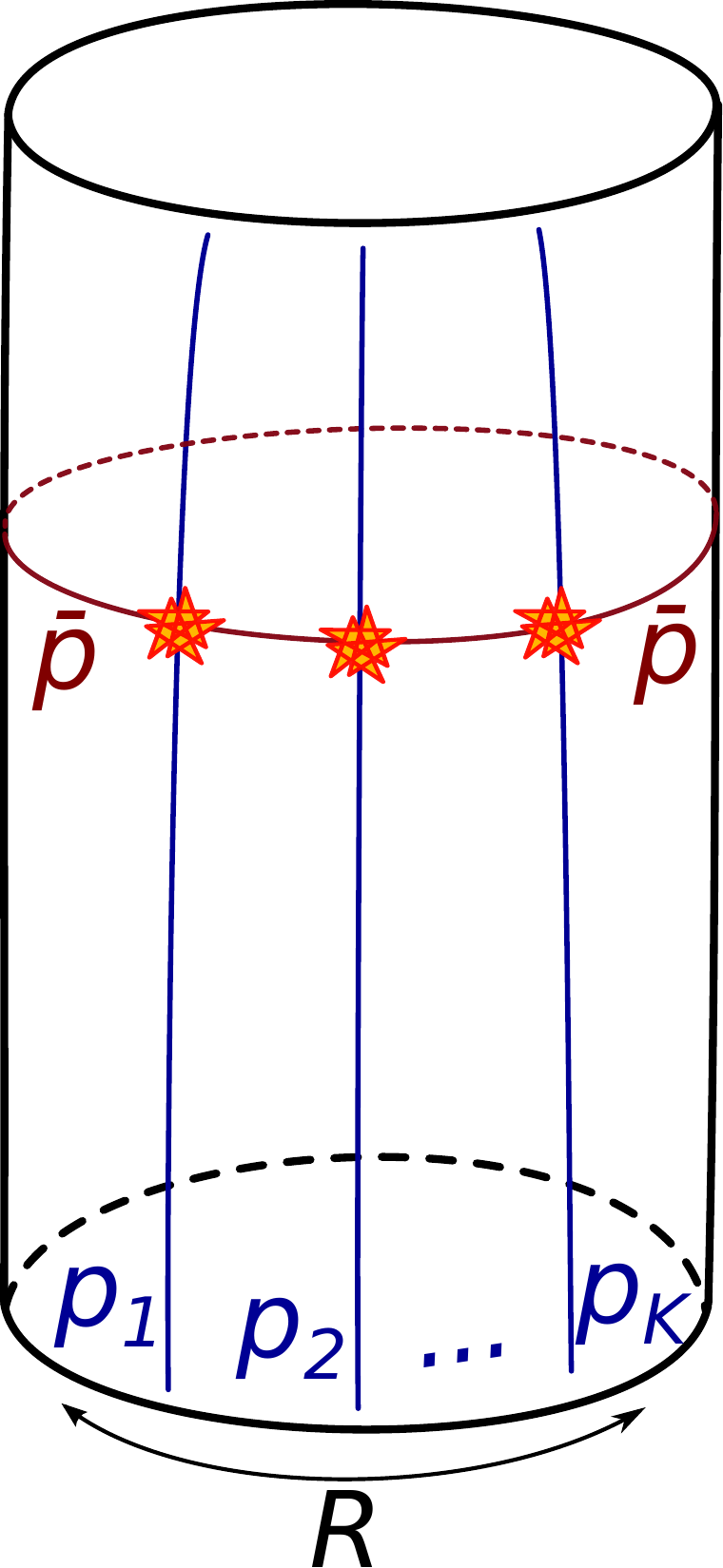}
  \caption{Pictorial representation of a wrapping effect. The worldsheet is a cylinder of size~$R$. One mirror particle of mirror momentum~$\bar{p}(u)$ (dark red) wraps around it, scattering in sequence with particles $1,2\dots K$ (blue). We should sum over all possible mirror particles and integrate over their rapidities~$u$.}
  \label{fig:wrapping}
\end{figure}
In this formula we integrate over a suitable rapidity $u$, $\bar{p}(u)$ is the ``mirror'' momentum, $\bar{H}(u)$ the mirror energy, $X$ denotes any possible virtual particle, and $(-1)^{F_X}$ its fermion sign. This process is pictorially represented in figure~\ref{fig:wrapping}.%
\footnote{%
Notice that the integrand~\eqref{eq:fterm} takes a simpler form with respect to the general expression for the F~terms, which is given by a transfer matrix. This is because we used the fact that here scattering is diagonal.}
 Specialising this formula to our case~\eqref{eq:Smatrix} and to the length~$R_0$ we get
\begin{equation}
\int \de u \frac{\partial \bar{p}(u)}{\partial u} e^{i \bar{H}(u)\, R_0}e^{i[\Phi(p(u),p_1)+\Phi(p(u),p_2)+\cdots +\Phi(p(u),p_K)]}\sum_{X}(-1)^{F_X}=0\,,
\end{equation}
where in the last step we used that virtual excitations also come in supersymmetric pairs---or rather, $(\mathbf{2}|\mathbf{2})$ multiplets. Hence, the integrand of the leading wrapping correction vanishes identically. This is exactly the same mechanism by which BPS states are protected against wrapping~\cite{Baggio:2017kza}, and extends to all finite-size effects. Notice that this argument uses the structure of the S~matrix~\eqref{eq:Smatrix} but is insensitive to the explicit form of the phase shift~$\Phi(p_i,p_j)$.
In fact, more can be done by taking into account the particularly simple form of the phase shift, $\Phi(p_i,p_j)\sim p_i\,p_j$. As for flat bosonic string theory which has a similar dynamics~\cite{Dubovsky:2012wk}, the factorised form of the phase shift leads to drastic simplifications of the (mirror) TBA equations. As a result, these can be solved exactly---which is hardly ever the case---leading to a simple formula for the finite-size energy. Indeed in this way we can prove that \textit{all wrapping effects cancel} and the mirror TBA equations coincide with the Bethe-Yang ones~\eqref{eq:BetheYang}, hence substantiating our claim that strings on pure-NS-NS $\AdSST$ are equivalent to a spin-chain with no wrapping effects.
While this is a bit tangential to the the main purpose of our paper, it is instructive to perform the mTBA construction explicitly as we do in the next section. 
However, readers that are so inclined can jump ahead to section~\ref{sec:spinchain} where we present the integrable spin-chain in a self-contained~way.

\section{Mirror thermodynamic Bethe ansatz}
\label{sec:mTBA}
In this section we discuss the mirror model of  pure-NS-NS $\AdSST$ strings and show that their mirror TBA equations coincide with the Bethe-Yang ones.
The idea is to introduce an auxiliary theory, the \emph{mirror model}, where the notions of space and time are exchanged. If our two-dimensional integrable model is {non-relativistic}, as in the case at hand and in general in AdS/CFT integrability, the mirror model is a genuinely new theory~\cite{Arutyunov:2007tc}. Still, as the original theory and its mirror are related by analytic continuation, the dynamics and indeed integrability of the mirror theory follow from that of the original~one.

\subsection{Mirror model}
\label{sec:mirrortheory}
Following the seminal paper ref.~\cite{Arutyunov:2007tc} we start by considering our integrable QFT in finite volume~$R$ and at finite temperature~$T=1/\beta$.%
\footnote{%
Here we work with the string-frame S~matrix of appendix~\ref{app:Smatframe}, so that the worldsheet length is the R~charge of each given state $R=j+\tilde{\jmath}$, rather than the R~charge of the vacuum~$R_0$. Our discussion is largely insensitive to these details.
}
 We are interested in the energy spectrum as~$R$ is finite and $\beta\to\infty$.
 The key step is to notice that the partition function of this theory coincides with that of a \textit{mirror theory} in finite-volume $T$ and at temperature $1/R$. Schematically
\begin{equation}
\label{eq:partition}
Z(\beta,L)= \text{Tr}\left[ e^{-\beta\, \Hgen}\right]=\text{Tr}\left[ e^{-R\, \bar{\Hgen}}\right]\,,
\end{equation}
where the mirror Hamiltonian is denoted by~$\bar{\Hgen}$.%
\footnote{We will denote all quantities in the mirror kinematics with bars, while reserving tildes to indicate right-movers on the worldsheet.}
Hence the zero-temperature, finite-volume original theory can be understood as the finite-temperature, infinite-volume mirror theory where the notions of time and space have been exchanged. As it can be seen by path-integral manipulations~\cite{Arutyunov:2007tc}, the mirror theory is related to the original one by two Wick rotations on the worldsheet, exchanging time and space. 
As a consequence, energy and momentum on the worldsheet transform as%
\footnote{In a sense, this is ``half'' of a crossing transformation $(H,p)\to(-H,-P)$. This can often be made precise by introducing appropriate rapidity variables under which the mirror transformation is a shift.}
\begin{equation}
\label{eq:stringmirrorHP}
H\to i \bar{p},\qquad p\to i \bar{H},
\end{equation}
so that for non-relativistic theories the original theory and its mirror do not coincide. Indeed performing the mirror transformation of~\eqref{eq:dispersion} yields
\begin{equation}
\label{eq:mirrorH}
\bar{H}(\bar{p},\mu)=\frac{2\pi}{k}\big(|\bar{p}|+i \mu\big)\,.
\end{equation}
It is a little concerning that the Hamiltonian features an imaginary shift. As we shall see below, $\mu$ should be interpreted as a chemical potential in the thermal partition function of the mirror model, rather than as part of the mirror energy.
We will assume that the S~matrix of the mirror theory is related to the original one by analytic continuation,
\begin{equation}
\label{eq:mirrorS}
\bar{S}(\bar{p}_1,\bar{p}_2)= S(p_1,p_2)\big|_{p_1=\bar{p_1},p_2=\bar{p_2}}\,.
\end{equation}
We can hence write down the Bethe-Yang equations for the mirror model,
\begin{equation}
e^{i \bar{p}_j \beta}\prod_{k\neq j}\bar{S}_{jk}(\bar{p}_j,\bar{p}_k)=(-1)^F\,.
\end{equation}
The fermion sign on the right-hand side takes into account that, while the fermions of the original theory were periodic, those of the mirror theory are anti-periodic~\cite{Arutyunov:2007tc}.
Starting from here, we will be able to compute the partition function of the mirror theory in large-volume limit $\beta\to\infty$ and at finite temperature~$1/R$, which in turn will give us the ground-state energy of the original theory in finite volume~$R$. Moreover, the spectrum of excited states can be similarly obtained~\cite{Dorey:1996re}.

\subsection{Mirror TBA and its solution}
The derivation of the mirror thermodynamic Bethe ansatz equations for a model with diagonal scattering such as our is rather well~known in the literature, see \textit{e.g.}~\cite{Bajnok:2010ke}. For the reader's convenience and to fix our conventions, we have collected it in appendix~\ref{app:mTBA}. The final result is a set of non-linear integral equations for the pseudo-energies~$\epsilon_a(u)$, where $a$ denotes the particle type and $u$ is a convenient rapidity variable. These read
\begin{equation}
\label{eq:TBAeq}
\epsilon_a(u)= \psi_a+R\, \bar{H}_a(u)-[\Lambda_b*K_{ba}](u)\,,\quad
\Lambda_b(v)=\begin{cases}
-\log(1-e^{-\epsilon_b(v)})& \text{bosons},\\
+\log(1+e^{-\epsilon_b(v)})& \text{fermions}.
\end{cases}
\end{equation}
where the kernel $K_{ab}$ is the logarithmic derivative of the mirror scattering matrix, see eqs.~(\ref{eq:deltaphase},~\ref{eq:Kkernel}), and $\psi_a$ is a chemical potential for the fermions
\begin{equation}
\psi_a=\begin{cases}
0&\text{bosons},\\
i\pi&\text{fermions},
\end{cases}
\end{equation}
see also eq.~\eqref{eq:psi-a}. One should solve this system of equations---which generally is only possible numerically---and use the resulting value of the pseudo-energies to compute the ground-state energy \textit{of the original model in finite volume}, 
\begin{equation}
\label{eq:TBAenergy}
H_0 = -\frac{1}{2\pi}\, (\partial_u\bar{p}_a)* \Lambda_a\,.
\end{equation}

We now want to solve eqs.~(\ref{eq:TBAeq}--\ref{eq:TBAenergy}) using the mirror kinematics described in section~\ref{sec:mirrortheory}. Given the form of the mirror dispersion~\eqref{eq:mirrorH} and S~matrix~\eqref{eq:mirrorS} it is convenient to take as ``rapidity'' $u$ the mirror momentum itself. Then it is convenient to distinguish again chiral and anti-chiral excitations on the (mirror) worldsheet. Unlike what happens in the original model, here chiral particles are exactly those of positive (mirror) momentum, and anti-chiral ones are those of negative momentum, as dictated by eq.~\eqref{eq:mirrorH}. Writing $\Phi(\bar p_b, \bar p_a)$ as 
\begin{equation}
\Phi(\bar{p}_b, \bar{p}_{a}) =
+\dfrac{k}{2\pi} \bar{H}(\bar{p}_b)\,\bar{H}(\bar{p}_a)\,\Theta(-\bar{p}_b)\Theta(\bar{p}_a) - \dfrac{k}{2\pi} \bar{H}(\bar{p}_b)\,\bar{H}(\bar{p}_a)\, \Theta(\bar{p}_b)\Theta(-\bar{p}_a) \,,
\end{equation}
we have
\begin{multline}
K_{ba}(\bar{p}_b,\bar{p}_a) =- \frac{1}{2\pi} \Big[\Theta(- \bar{p}_b) + i\mu_b\,\delta(\bar{p}_b)\Big]\,\bar{H}(\bar{p}_a)\,\Theta(\bar{p}_a)\\
 - \frac{1}{2\pi} \Big[\Theta(\bar{p}_b) +i\mu_b\,\delta(\bar{p}_b)\Big]\,\bar{H}(\bar{p}_a)\,\Theta(- \bar{p}_a) \,.
\end{multline}
Splitting the TBA equation between left and right movers, and denoting the latter with tildes we have
\begin{equation}
\label{eq:TBAexplicit}
\begin{aligned}
\epsilon_a(\bar{p})=\psi_a+ \bar H(\bar p) \left[ R + \frac{1}{2 \pi}\,\sum_{b}\int\limits_{-\infty}^{0}\de\bar{p}'\tilde{\Lambda}_b(\bar{p}') - \dfrac{i}{2 \pi} \sum_b \mu_b \, \Lambda_b (0) \right] \,,\\
\tilde{\epsilon}_a(\bar{p})=\psi_a + \bar H(\bar p) \left[ R + \frac{1}{2 \pi}\,\sum_{b}\int\limits_{0}^{+ \infty}\de\bar{p}'\tilde{\Lambda}_b(\bar{p}') - \dfrac{i}{2 \pi} \sum_b \mu_b \, \Lambda_b (0) \right] \,.
\end{aligned}
\end{equation}
Indeed as anticipated $i\mu$ acts as a chemical potential in the mirror theory.
By taking two derivatives with respect to~$\bar{p}$ we see that the pseudo-energy $\epsilon_a$ and $\tilde{\epsilon}_a$ are at most (affine) \textit{linear} functions of~$\bar{p}$. Furthermore, $\epsilon_a$ (or $\tilde{\epsilon}_a$) does not really depend on the particle flavour~$a$, but only on $\mu$ and on whether the excitation is bosonic or fermionic, due to the presence of $\psi_a$. This leads to the following ansatz for bosonic and fermionic pseudo-energies:
\begin{equation}
\label{eq:ansatzc}
\begin{aligned}
\epsilon_a(\bar{p})&=\ c_1(\mu)\,\bar{p}+c_0(\mu),\qquad &&a\sim\text{boson}\,,\\
\epsilon_a(\bar{p})&=\ c_1(\mu)\,\bar{p}+c_0(\mu)+i\pi,\qquad &&a\sim\text{fermion}\,,
\end{aligned}
\end{equation}
and similarly for~$\tilde{\epsilon}_a$, depending on~$\tilde{c}_1,\tilde{c}_0$.
Using the fact that for each value of~$\mu$ we have $({\bf 2}|{\bf 2})$ bosons and fermions and the explicit form of~$\Lambda_a$ from eq.~\eqref{eq:TBAeq}, we get
\begin{equation}
\begin{aligned}
\sum_b \int \de \bar{p}\, \Lambda_b(\bar{p}) & = \int \de \bar{p}\, \sum_{\mu}2\left[-\log(1-e^{-\epsilon(\bar{p},\mu)})+\log(1+e^{-\epsilon(\bar{p},\mu)+i\pi})\right]=0\,, \\
\sum_b \mu_b \, \Lambda_b (0) & = \sum_{\mu}2 \mu \left[-\log(1-e^{-\epsilon(\bar{p},\mu)})+\log(1+e^{-\epsilon(\bar{p},\mu)+i\pi})\right]=0\,,
\end{aligned}
\end{equation}
and similarly for~$\tilde{\Lambda}_b*K_{ba}$. As the integrands vanish, the solution for~\eqref{eq:ansatzc} can be easily read off eq.~\eqref{eq:TBAexplicit},
\begin{equation}
c_1(\mu)=-\tilde{c}_1(\mu)=\frac{2\pi R}{k},\qquad c_0(\mu)=\tilde{c}_0(\mu)=\frac{2\pi i R}{k}\mu\,.
\end{equation}
 Regardless of the explicit value of these coefficients, by the same cancellation mechanism the integrand of the ground-state energy~$H_0$ vanishes, so that
\begin{equation}
H_0=0\,,
\end{equation}
as dictated by supersymmetry. This of course needed to be the case, and is unremarkable in itself. However, we shall see below that the non-trivial part of the pseudo-energies cancels \textit{even for excited states}, proving the triviality of wrapping corrections.

\subsection{Excited states}
Once the ground-state mirror TBA equations are known, the equations for excited states can be derived by the \textit{contour-deformation} approach of Dorey and Tateo~\cite{Dorey:1996re}. The idea is that excited states are given by the same TBA equations, up to taking the integration contour to be some appropriate curve. Of course as long as the contour is close to the vacuum one (the real mirror-momentum line) we do not find anything new, as we can deform it back to the real line. New equations do arise, however, if in deforming the contour we encounter a singularity of the TBA integrands. Given the form of~$\Lambda_b(u)$ in eq.~\eqref{eq:TBAeq}, this can happen if
\begin{equation}
\label{eq:TBAsing}
e^{-\epsilon_b(u)}|_{u=u^*}=+1\text{ (bosons)},\qquad
e^{-\epsilon_b(u)}|_{u=u^*}=-1\text{ (fermions)}.
\end{equation}
Then, $\Lambda_b(u^*)\sim\log0$. Integrating by parts and using Cauchy's theorem, in the TBA equations~\eqref{eq:TBAeq} we pick up a term of the form~$i\varphi_{ba}(u^*,u)$, see eq.~\eqref{eq:deltaphase}. This happens for each of the $\{u^*_j\}$ satisfying~\eqref{eq:TBAsing}; moreover, for bosons we can encircle the same $u^*_j$ multiple times, much like in ref.~\cite{Dubovsky:2012wk}.
All in all, we obtain the following mirror TBA equations:
\begin{equation}
\label{eq:TBAdriving}
\epsilon_a(u)= \psi_a+R\, \bar{H}_a(u)-[\Lambda_b*K_{ba}](u)+i\sum_j \Phi_{ba}(u^*_j,u)\,,
\end{equation}
subject to the condition~\eqref{eq:TBAsing} which we can write as
\begin{equation}
\label{eq:TBAquant}
-\epsilon_{a_j}(u^*_j)=2\pi i \nu_j-\psi_{a_j}\,,\qquad
\nu_j\in\mathbb{Z}\,.
\end{equation}
In the formula for the energy~\eqref{eq:TBAenergy}, we similarly pick up some residues:
\begin{equation}
\label{eq:TBAHdriving}
\Htot=-\frac{1}{2\pi}(\partial_u \bar{p}_a)*\Lambda_a+i \sum_j \bar{p}(u^*_j)\,.
\end{equation}

We can repeat almost \textit{verbatim} the arguments of the previous subsection to find that the integrand in eq.~\eqref{eq:TBAdriving} vanishes, and that the right-hand side of the mirror TBA equations does not depend on the pseudoenergies. The quantisation condition~\eqref{eq:TBAquant} is then
\begin{equation}
R\, \bar{H}_a(u^*_j)-i\sum_{k}\Phi_{b_ka_j}(u^*_k,u^*_j)=-2\pi i \nu_j\,.
\end{equation}
Recalling that $\bar{H}=-ip$~\eqref{eq:stringmirrorHP}, and that $\Phi_{ab}=-i\log S_{ab}$ \eqref{eq:deltaphase} is anti-symmetric by unitarity, we obtain
\begin{equation}
iR\, p_j+\sum_k \log S_{jk}(p_j,p_k)=2\pi i \nu_j\,,
\end{equation}
which is nothing but the Bethe-Yang equations of the original model. From eq.~\eqref{eq:stringmirrorHP} we also observe that $\bar{p}=-i H$, so that eq.~\eqref{eq:TBAHdriving} becomes
\begin{equation}
\Htot=\sum_j H(p_j)\,.
\end{equation}
This proves that~\textit{the mirror TBA equations coincide with the Bethe-Yang equations} for this model.

\section{Proposal for an integrable spin~chain}
\label{sec:spinchain}
Motivated by the observations of the previous section, we propose an integrable spin~chain describing the spectrum of closed strings on $\AdSST$ with pure-NS-NS background fluxes.

\subsection{Definition of the spin-chain}
We take a spin chain of integer length~$R_0$, with $(8|8)$ magnons with quantum numbers as in table~\ref{tab:excitations}, each carrying energy
\begin{equation}
\label{eq:scdispersion}
H(p,\mu) = \left|\frac{k}{2\pi}p+\mu\right|\,.
\end{equation}
The exact S~matrix is proportional to the identity and is given by
\begin{equation}
\label{eq:phidef}
S_{jk}=e^{i \Phi_{jk}},\qquad
\Phi_{jk}=\dfrac{1}{2} \left( p_jH_k-p_k H_j- p_j m_k+p_k m_j \right)\,,
\end{equation}
where we introduced~$m_j$ given by
\begin{equation}
\label{eq:mdefinition}
m_j = \mu_j\,\text{sgn}\big(\frac{k}{2\pi}p_j+\mu_j\big)\,.
\end{equation}
Distinguishing magnons by the sign of their velocity, $\partial_p H = \pm k/2\pi$, we can work out the explicit form of the S~matrix. This does not depend on $\mu$ and takes the form
\begin{equation}
S^{\pm\pm}(p_i,p_j)=1\,,\qquad S^{\mp\pm}(p_i,p_j)= \exp\left[\pm \frac{ik}{2\pi}\, p_i\,p_j\right]\,.
\end{equation}
The Bethe ansatz equations read
\begin{equation}
\label{eq:betheeqs}
e^{i p_i R_0}\prod_j e^{i \Phi_{ij}}=1\,.
\end{equation}
They can also be split depending on the particles' velocities,
\begin{equation}
\label{eq:BAE}
\begin{cases}
\displaystyle
e^{i p_i R_0}\prod_{j\text{ right}}S^{+-}(p_i,p_j)=1,\qquad& p_i>-\frac{2\pi\mu}{k}\quad\text{(chiral)}\,,\\
\displaystyle
e^{i p_i R_0}\prod_{\,\,j\text{ left}~}S^{-+}(p_i,p_j)=1,\qquad& p_i<-\frac{2\pi\mu}{k}\quad\text{(anti-chiral)}\,,
\end{cases}
\end{equation}
and are supplemented by the cyclicity constraint
\begin{equation}
1=\exp[i\sum_j p_j]=\exp[i\Ptot]=\exp[iP+i\tilde{P}]\,,
\end{equation}
where we have split the total momentum in its chiral and anti-chiral parts depending on the sign of~$\partial_p H$. The spectrum is therefore divided in superselection sectors with
\begin{equation}
\label{eq:ciclicity}
P+\tilde{P}=2\pi W,\qquad W\in\mathbb{Z}\,.
\end{equation}
Finally, the total energy of a state is
\begin{equation}
\Htot =\sum_{j} H(p_j,\mu_j)= H +\tilde{H}\,.
\end{equation}
which we also have similarly split.

\subsection{Solution of the Bethe ansatz}
Unlike what happens for most integrable spin~chains, here it is possible to solve explicitly the Bethe ansatz equations. We first shall do that in the sector where $W=0$, where these are a little simpler.

\paragraph{States without ``winding''.}
When $\Ptot=0$ we take the logarithm of the Bethe equations~\eqref{eq:betheeqs} and find
\begin{equation}
p_i R_0 + \sum_{j}\Phi_{ij}=
p_i \big[R_0+\frac{1}{2}\sum_{j}(H_j-m_j)\big]=2\pi \nu_i\,,
\end{equation}
where in the first equality we used eq.~\eqref{eq:phidef} and the vanishing of the total momentum to simplify $\sum_{j}\Phi_{ij}$.
The Bethe equations can be expressed in terms of $\Htot$ and~$\Mtot$,
\begin{equation}
\label{eq:Betheeqlinear}
p_i \big(R_0+\frac{\Htot-\Mtot}{2}\big)=2\pi \nu_i\,,
\end{equation}
so that we can establish a linear relation between momenta~$p_i$ and levels~$\nu_i$,
\begin{equation}
\label{eq:scindividualquant}
p_i=\frac{2\pi}{\Reff}\nu_i,\qquad
\Reff=R_0+\frac{\Htot-\Mtot}{2}\,.
\end{equation}
Plugging these values into the dispersion~\eqref{eq:scdispersion}, we have
\begin{equation}
\label{eq:scenergysol}
\Htot= \sum_i \left|\frac{k}{\Reff}\nu_i+\mu_i\right|\,.
\end{equation}
In order to proceed and solve this equation we need to remove the absolute value depending on the value of each $p_i= 2\pi \nu_i/\Reff$ with respect to $\mu_i$. For the moment, we do so implicitly, using $P=2\pi \Nsc/\Reff$ for chiral excitations and $\tilde{P}=-2\pi \tilde{\Nsc}/\Reff$ for anti-chiral ones. Here $\Nsc=\sum_{\text{chiral}}\nu_i$, and similarly~$\tilde{\Nsc}$. Notice that then
\begin{equation}
\Ptot=P+\tilde{P}=0\qquad\Leftrightarrow\qquad
\Nsc=\tilde{\Nsc}\,,
\end{equation}
and the total energy follows from eq.~\eqref{eq:scenergysol}
\begin{equation}
\Htot = \frac{2k}{\Reff}\Nsc+\Mtot,\qquad
\Mtot =\sum_{j} m(p_j,\mu_j)= M +\tilde{M}\,. 
\end{equation}
Using the definition of $\Reff$~\eqref{eq:scindividualquant} we find 
\begin{equation}
\label{eq:SpinChainSpectrum-W0}
\Htot= \sqrt{R_0^2+4k\,\Nsc}-R_0+M+\bar{M}\,.
\end{equation}

\paragraph{General solution.}
In the general $W\neq 0$ case it is more convenient to split the Bethe equations according to the particles' chiralities, like in eq.~\eqref{eq:BAE}.
Let us also observe that the S~matrix satisfies
\begin{equation}
-i\log S^{\mp\pm}(p_i,p_j)=p_i (H_j\mp\mu_j)= p_i (H_j-m_j)\,,
\end{equation}
where $m_i$ is given in eq.~\eqref{eq:mdefinition}. Furthermore,
\begin{equation}
\label{eq:prel}
H=\frac{k}{2\pi}P + M\,,\qquad
\tilde{H}=-\frac{k}{2\pi}\tilde{P} + \tilde{M}\,.
\end{equation}
Using this notation, we can take the logarithm of the Bethe equations~\eqref{eq:BAE}
\begin{equation}
\begin{cases}
\displaystyle
p_i (R_0+\tilde{H}-\tilde{M})=+2\pi \nu_i,\qquad& p_i>-\frac{2\pi\mu}{k}\quad\text{(left)}\,,\\
\displaystyle
\tilde{p}_i (R_0+H-M)=-2\pi \tilde{\nu}_i,\qquad& \tilde{p}_i<-\frac{2\pi\mu}{k}\quad\text{(right)}\,.
\end{cases}
\end{equation}
Notice that $p_i$ is generally positive (resp.~negative) for chiral (resp.~anti-chiral) magnons. Since $R_0+H-M$, $R_0+\tilde{H}-\tilde{M}$ are always positive, we have explicitly picked the sign of the integers $\nu_i$, $\tilde{\nu}_i$.
Summing over ``left'' and ``right'' particles, we find
\begin{equation}
\begin{cases}
\displaystyle
+\frac{2\pi}{k}(H-M) (R_0+\tilde{H}-\tilde{M})=+2\pi \Nsc,\\[0.2cm]
\displaystyle
-\frac{2\pi}{k}(\tilde{H}-\tilde{M}) (R_0+H-M)=-2\pi \widetilde{\Nsc}.
\end{cases}
\end{equation}
The Bethe equations can be readily solved for the total energy $\Htot=H+\tilde{H}$ and for the total momentum $\Ptot=P+\tilde{P}$
\begin{equation}
\label{eq:SpinChainSpectrum}
\Htot= \sqrt{R_0^2+2k (\Nsc+\tilde{\Nsc})+k^2W^2}-R_0+\Mtot\,,\qquad
\Ptot=\frac{2\pi}{R_0}(\Nsc-\tilde{\Nsc})=2\pi W\,,
\end{equation}
where we used eq.~\eqref{eq:prel} and the ciclicity constraint~\eqref{eq:ciclicity}.
Notice that when $W=0$ the solution reduces to eq.~\eqref{eq:SpinChainSpectrum-W0}.


\subsection{Identification with the string quantum numbers}
\label{sec:spinchaincomparison}
In the previous subsection we have introduced the spin-chain length~$R_0$, the energy $H$ and the ``mass'' $M$ without any reference to the dual string theory. However, based on the discussion of appendix~\ref{app:Smatframe} we expect that $R_0$ is the R~charge of a BPS vacuum state~$|\Omega_{R_0}\rangle$,
\begin{equation}
\label{eq:vacuum}
\big(\Jgen^3+\tilde{\Jgen}^3\big)\,|\Omega_{R_0}\rangle = R_0\,|\Omega_{R_0}\rangle\,,
\end{equation}
while $\Htot$ gives the \textit{light-cone energy}~\eqref{eq:lcHam} of an excitation. Similarly $M$ and $\tilde{M}$ are the contributions due to the charges of the chiral and anti-chiral oscillators,%
\footnote{%
\label{ft:signs}
We have implicitly defined $\delta j$ and $\delta \tilde{\jmath}$ with a minus sing is such a way that they contribute positively to $M, \tilde{M}$ and hence to the Hamiltonian; this is merely a matter of convenience.}
\begin{equation}
M = \sum_i\big(\delta \ell_i+ \delta j_i\big),\qquad
\tilde{M}=\sum_i\big(\delta \tilde{\ell}_i+\delta \tilde{\jmath}_i\big)\,,
\end{equation}
where $\delta \ell_i, \delta j_i$ are the zero-momentum contribution to the (left) energy and R~charge due to the $i$-th particle, \textit{cf.}~table~\ref{tab:excitationszeromom}. Notice that indeed for the vacuum $\Htot=0$ as required by the BPS condition.

\begin{table}[t]
\centering
\renewcommand{\arraystretch}{1.2}
\begin{tabular}{|c|c|c|c|}
\hline
Chiral part. & $m=\mu$ & $\delta\ell$ & $\delta j$ \\
\hline 
$Y(p)$ & $1$ & $0$ & $1$\\
$\eta^a(p)$ & $1$ & $\tfrac{1}{2}$& $ \tfrac{1}{2}$ \\
$Z(p)$ & $1$ & $1$ & 0  \\
\hline
$\bar Z(p)$ & $-1$ & $-1$& 0 \\
$\bar \eta^a(p)$ & $-1$& $- \tfrac{1}{2}$ & $- \tfrac{1}{2}$ \\
$\bar Y(p)$ & $-1$ & $0$ & $-1$   \\
\hline
$\chi^{\dot{a}}(p)$ & $0$& $-\tfrac{1}{2}$ & $\tfrac{1}{2}$\\
$T^{a\dot{a}}(p)$ & $0$& $0$ & $0$\\
$\bar{\chi}^{\dot{a}}(p)$ & $0$& $\tfrac{1}{2}$ & $-\tfrac{1}{2}$\\
\hline
\end{tabular}
\hspace*{1cm}\begin{tabular}{|c|c|c|c|}
\hline
Anti-chiral part. & $\tilde{m}=-\mu$ & $\delta\tilde{\ell}$ & $\delta\tilde{\jmath}$ \\
\hline 
$Y(p)$ & $-1$ & $0$ & $-1$\\
$\eta^a(p)$ & $-1$ & $-\tfrac{1}{2}$& $ -\tfrac{1}{2}$ \\
$Z(p)$ & $-1$ & $-1$ & $0$  \\
\hline
$\bar Z(p)$ & $1$ & $1$& $0$ \\
$\bar \eta^a(p)$ & $1$& $\tfrac{1}{2}$ & $ \tfrac{1}{2}$ \\
$\bar Y(p)$ & $1$ & $0$ & $1$   \\
\hline
$\chi^{\dot{a}}(p)$ & $0$& $\tfrac{1}{2}$ & $-\tfrac{1}{2}$\\
$T^{a\dot{a}}(p)$ & $0$& $0$ & $0$\\
$\bar{\chi}^{\dot{a}}(p)$ & $0$& $-\tfrac{1}{2}$ & $\tfrac{1}{2}$\\
\hline
\end{tabular}
\renewcommand{\arraystretch}{1}
\caption{%
In the left table, we list contributions to $\delta \ell$, $\delta j$ and $m=\delta\ell+\delta j$ for chiral magnons; these have $\delta \tilde{\ell}=\delta \tilde{\jmath}=\tilde{m}=0$. Similarly, in the right table we list the charges of the anti-chiral magnons. This table follows from table~\ref{tab:excitations} by identifying the charge shifts~$\delta \ell$, \textit{etc.}\ with the charges of a string excitation ``at rest'', \textit{i.e.}\ with $p=0$.
}
\label{tab:excitationszeromom}
\end{table}

\label{sec:sccomments}
\paragraph{Zero-energy (BPS) states.}
It is interesting to notice that the vacuum is not the only state with $\Htot=0$. In fact we have four bosonic and four fermionic excitations with $H(p)=k|p|/2\pi$. Clearly, adding one such excitation to the vacuum with $p=0$ yields another state of zero energy. What is more, it is easy to check that such a state always solves the Bethe equations. This is a little disconcerting for zero-momentum bosons because they seemingly  generate infinitely many zero-energy states. This signals some additional symmetry of the model, which becomes clear by thinking of the original $\AdSST$ description: those four bosons correspond to excitations of the $\T^4$ directions, which are flat and hence have a $\u(1)^{\oplus 4}$ symmetry. The bosonic zero-modes correspond to those shifts, and we hence exclude them from the spectrum. It is more interesting to consider the fermionic zero modes, which instead generate a finite number of states---16, to be precise. As discussed in ref.~\cite{Baggio:2017kza}, these reproduce the spectrum of BPS states of the model, \textit{cf.}\ eq.~\eqref{eq:HodgeDiamond}.

\paragraph{WZW interpretation and two puzzles.}
Using the above identifications we obtain from eq.~\eqref{eq:SpinChainSpectrum-W0}
\begin{equation}
\Htot= 
\sqrt{R_0^2+4k\, \Nsc}-R_0+\delta \ell+\delta j+\delta \tilde{\ell}+\delta \tilde{\jmath}\,,
\end{equation} 
which is valid in the simplest sector with $W=0$.%
\footnote{Notice the sign of $\delta j, \delta \tilde{\jmath}$, see also footnote~\ref{ft:signs}.}
Running a little ahead of ourselves, we notice that this formula bears a striking resemblance with the solution of the mass-shell condition of the $\AdSST$ WZW model~\cite{Baggio:2018gct}, see appendix~\ref{app:WZW} and in particular eq.~\eqref{eq:unflowed-mass-shell-solution}, provided that we identify the (chiral) spin-chain excitation number $\Nsc$ with the total (chiral) WZW excitation number~$\NN$.
This however raises two related issues. Firstly, $\NN=\NNR$ in the WZW model, whereas in the spin chain $\Nsc\neq\tilde{\Nsc}$ when $W\neq 0$. Secondly but perhaps more importantly, in the WZW model $\NN=\sum_i n_i\geq 0$ and indeed $n_i\geq 0$ for a physical state. On the other hand here we have, for chiral spin-chain excitations,
\begin{equation}
\nu_i> - \frac{\Reff}{k}\,\mu_i\,,
\end{equation}
which is a non-negative number only if
\begin{equation}
k>\Reff=R_0+\frac{\Htot-\Mtot}{2}\,.
\end{equation}
However, there is no reason to assume that this is the case in the spin chain, as $\Reff$ and indeed $R_0$ are not bounded. Clearly, more care is needed in identifying $\nu_i$ with $n_i$ and hence $\Nsc$ with~$\NN$.

The course of this confusion is that, unlike the energy~$\ell$ and R~charge~$j$, the excitation numbers~$\nu_i$ and $n_i$ are not observables, but rather internal labels of our description. This makes their identification less straightforward.
We will see in detail in the next section that such a matching is indeed possible, and it will require distinguishing between different \textit{spectrally flowed} sectors, see also appendix~\ref{app:WZW}. Taking this into account, we shall see that the Bethe equations~\eqref{eq:BAE} perfectly reproduce the mass-shell condition of the Wess-Zumino-Witten model.

\section{WZW spectrum and comparison}
\label{sec:WZWcomp}
The WZW description of strings propagating on $\AdSST$ with pure-NS-NS flux is well known~\cite{Giveon:1998ns,Maldacena:2000hw,Pakman:2003cu,Israel:2003ry,Raju:2007uj,Giribet:2007wp, Ferreira:2017pgt} and it is briefly reviewed in appendix~\ref{app:WZW} for the reader's convenience. Physical states should be annihilated by the non-negative modes of the super-Virasoro algebra on the worldsheet. This leads to a mass-shell condition, \textit{i.e.}\ a quadratic constraint, resulting in a square-root formula for the light-cone energy, as reviewed in the appendix. 
We claim that this constraint is equivalent to the one imposed by the Bethe equations of the spin~chain of section~\ref{sec:spinchain}, once  suitable identifications have been made. These are rather straightforward and present only two obstacles.
Firstly, when comparing (RNS) worldsheet fermions to fermonic degrees of freedom in target space (in our case, in the dual spin~chain), the usual subtleties arise and it is necessary to properly account for the GSO projection. Secondly, as reviewed in appendix~\ref{app:WZW}, in the WZW language we should consider various spectrally-flowed sectors, labelled by an integer~$w$; this is indeed one of the parameters entering the mass-shell condition, and it makes the comparison a little more involved. Below we detail the matching of the spin-chain quantisation conditions with the WZW mass-shell condition starting from the simpler ``spectrally unflowed'' sector where $w=0$.

\subsection{The unflowed sector}
\label{sec:unflowed-comparison}

In the chiral sector, the WZW spectrum is determined in terms of two numbers $\ell_0$ and $j_0$, which are the weights of a lowest (highest) weight state of $\sl(2)_\L$ ($\su(2)_\L$), see appendix~\ref{app:WZW}. As usual, $j_0$ is half-integer, while $\ell_0$ is real. In the unflowed sector the mass-shell condition \eqref{eq:unflowed-mass-shell} gives, for the chiral sector,
\begin{equation}
\ell_0 = \frac{1}{2} + \frac{1}{2} \sqrt{(2 j_0 + 1)^2 + 4k \NN} \,,
\end{equation}
where $\NN$ is the total level of chiral excitations, \textit{cf.}\ eq.~\ref{eq:Neff}. We can think that $\ell_0$ and $j_0$ identify respectively the (left) energy and R~charge of a suitable vacuum state. The true (left) energy and R~charge of the state is given by eq.~\eqref{eq:unflowed-spins-NS} so that we find for the left contribution to the worldsheet Hamiltonian,%
\footnote{%
Notice that the sign of $\delta j$ and $s_j$ have been chosen so that they give a positive contribution to the light-cone Hamiltonian.
}
\begin{equation}
H=\ell - j = \ell_0 - j_0 + \delta \ell + \delta j
\end{equation}
in the NS sector and similarly, in the R sector
\begin{equation}
H=\ell - j = \ell_0 - j_0 + \delta \ell + \delta j + s_\ell + s_j \,. 
\end{equation}
Here $\delta \ell$ and $\delta j$ are the shifts in $\sl(2)_{\L}$ and $\su(2)_{\L}$ charge with respect to the vacuum ones $(\ell_0,j_0)$, due to the action of the Ka\v{c}-Moody modes. An additional shift appears in the R sector, where $s_\ell=\pm1/2$ and $s_j=\pm1/2$ identify the choice of fermionic ground-states.
Putting together left and right movers, and taking into account the level-matching condition~\eqref{eq:unflowed-level-matching} we find
\begin{equation}
\Htot = \sqrt{(2 j_0 + 1)^2 + 4k \NN} - (2j_0 + 1) + \delta+\tilde{\delta}\,,
\label{eq:WZW-unflowed-dispersion}
\end{equation} 
where we have defined
\begin{equation}
\label{eq:delta}
\delta = \delta \ell + \delta j  + s_\ell  + s_j + 1\,,
\qquad
\tilde{\delta} =  \delta \tilde \ell + \delta \tilde \jmath +  \tilde s_\ell + \tilde s_j + 1\,.
\end{equation}
With a small abuse of notation we use the same expression in the NS and R sector, with the understanding that in the NS sector $s_\ell=s_j=0$ in~$\delta$, and similarly for $\tilde{\delta}$.
As discussed in appendix~\ref{app:WZWBPS}, for BPS states $\delta = 0$, $\NN = 0$ and we find $\Htot=0$ as expected.  

\paragraph{Spin-chain ground state.} Let us now go back to the spin-chain description. According to our dictionary, $R_0$ is the charge of a BPS vacuum state~\eqref{eq:vacuum}. Here we can read off that
\begin{equation}
\label{eq:Rzero}
R_0= 2j_0+1\,.
\end{equation}
This indeed identifies one state in the middle of the Hodge diamond~\eqref{eq:HodgeDiamond} which sits in the R-R sector in the WZW description. More specifically, we should take
\begin{equation}
\label{eq:RsectorBPS}
s_\ell=s_j=-\frac{1}{2},\qquad
\tilde{s}_\ell=\tilde{s}_j=-\frac{1}{2},
\end{equation}
which identifies four states depending on the choice of the remaining fermionic ground states in the R-R sector. In the notation of eq.~\eqref{eq:HodgeDiamond} these states take the form $(j_0+\tfrac{1}{2},j_0+\tfrac{1}{2})^{\dot{a}\dot{b}}$, and sit in the $\bf 3\oplus \bf1$ representation of $\su(2)_\circ$ (see section~\ref{sec:symmetries}). It is natural to take as our vacuum the singlet
\begin{equation}
\label{eq:scvacuum}
|\Omega_{R_0}\rangle\ = \epsilon_{\dot a \dot b} \ket{j_0 + \tfrac{1}{2}}^{\dot a} \otimes \ket{j_0 + \tfrac{1}{2}}^{\dot b} \,. 
\end{equation}

\paragraph{Other BPS states.} In the spin-chain description a state featuring $K$ chiral excitations above the BPS vacuum carries a charge of
\begin{equation}
M=\sum_{i=1}^K\delta \ell_i+\delta j_i\,.
\end{equation}
This is zero for BPS states.
For consistency with the WZW description such a charge should match with~$\delta$ which is given by eq.~\eqref{eq:delta}. Remark that the two formulae differ by a finite shift, which is due to the fact that $M$ is measured with respect \textit{to the BPS (spin-chain) vacuum}, whereas in the WZW model $\delta$ is measured with respect to the NS- or R-sector vacuum. The latter is not necessarily a BPS state, and in fact in the NS sector is not even a physical state (due to the GSO projection, see appendix~\ref{app:WZWBPS}).
Bearing this in mind, the identification works for the BPS states of the R sector, due to eq.~\eqref{eq:RsectorBPS}, as well as for those in the NS sector that have $\delta=M=0$ as required, see again appendix~\ref{app:WZWBPS}. In the WZW description different BPS states arise from different R/NS sectors; in the spin-chain they emerge from acting with fermion zero-modes, much like  in the Green-Schwarz description~\cite{Baggio:2017kza}.

\renewcommand{\arraystretch}{1.2}
\begin{table}[t]
\centering
\begin{tabular}{|c|c|c|c|c|c|c|}
\hline
WZW & $\delta \ell$ & $\delta \tilde \ell$ & $\delta j$ & $\delta \tilde \jmath$ & Spin chain & $m+\tilde{m}$ \\
\hline
$L_{-n}^+$ & $+ 1$ &  &  &  & $Z (\frac{2 \pi}{\Reff}n )$ & $+ 1$ \\
$L_{-n}^-$ & $- 1$ &  &  &  & $\bar Z (\frac{2 \pi}{\Reff}n )$ & $- 1$ \\
$J_{-n}^+$ &  & & $- 1$ &  & $\bar Y (\frac{2 \pi}{\Reff}n )$ & $- 1$ \\
$J_{-n}^-$ &  & & $+ 1$ &  & $ Y (\frac{2 \pi}{\Reff}n )$ & +1 \\
\hline
$\tilde L_{-n}^+$ &  & $+ 1$ &  &  & $\bar Z (-\frac{2 \pi}{\Reff}n )$ & $+ 1$  \\
$\tilde L_{-n}^-$ &  & $- 1$ &  &  & $ Z (-\frac{2 \pi}{\Reff}n )$ & $- 1$  \\
$\tilde J_{-n}^+$ &  &  &  & $- 1$ & $Y (- \frac{2 \pi}{\Reff}n )$ & $- 1$  \\
$\tilde J_{-n}^-$ &  &  &  & $+ 1$ & $\bar Y (- \frac{2 \pi}{\Reff}n )$ & $+1$  \\
\hline
$L_0^+$ & +1 &  &  &  & $Z (0)$ & +1 \\
$\tilde L_0^+$ &  & +1 &  &  & $\bar Z (0)$ & +1 \\
$J_0^-$ &  &  & +1 &  & $Y (0 )$ & +1  \\
$\tilde J_0^-$ &  &  &  & +1 &  $\bar Y (0)$ & +1 \\
\hline
\end{tabular}
\caption{We match the bosonic excitations in the unflowed sector of the WZW model to the spin~chain, along with the respective mode numbers (or momenta). In the first box, we list the chiral excitations with $n\geq1$; in the second one, the anti-chiral excitations with~$\tilde{n}\geq1$. The zero-modes are listed separately, and discussed below. Notice that while in the chiral sector we have $\textit{e.g.}$\ $J^-\sim Y$, in the anti-chiral one we have~$\tilde{J}^+\sim \bar{Y}$, see also figure~\ref{fig:unflowed}.}
\label{tab:bosonsmatch}
\end{table}
\renewcommand{\arraystretch}{1}

\paragraph{Bound on the spin-chain length.} 
Using the value of the vacuum R~charge~\eqref{eq:Rzero} we can find a compact expression for the effective length~$\Reff$:
\begin{equation}
\Reff=R_0+\frac{\Htot-\Mtot}{2}=2j_0+1+\frac{2\ell_0-2j_0-2}{2}=j_0+\ell_0\,,
\end{equation}
where we used $\Mtot=\delta+\tilde{\delta}$. As long as we are in the spectrally unflowed sector, this quantity is bounded by the Maldacena-Ooguri bound and by the unitarity bound~\eqref{eq:spin-bounds},
\begin{equation}
\label{eq:Rbound}
\frac{1}{2}<\Reff < k-\frac{1}{2}\,.
\end{equation}
From the point of view of the spin-chain, while the lower bound is quite reasonable (indeed we would naturally require $R_0\geq1$), the upper bound might appear arbitrary. We shall see in a moment that this arises quite naturally in the spin chain too.

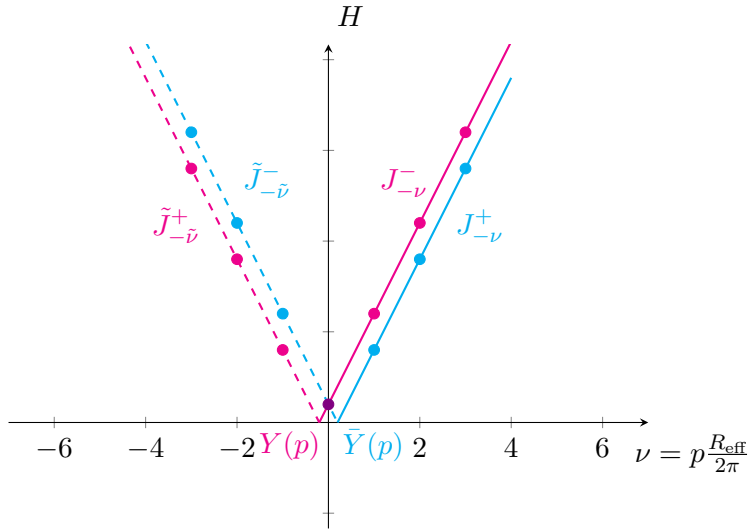
\begin{figure}
\centering
\begin{tikzpicture}
\begin{axis}[xlabel={},ylabel={},
height=8cm, width=10cm,
xmin=-7,xmax=7,ymin=-1,ymax=7, axis lines=center, yticklabels={,,}, axis equal]

\addplot[domain=-0.2:4, color=magenta,thick]{2*x+0.4};

\addplot[domain=-4.5:-0.2, color=magenta,thick, dashed]{-2*x-0.4};

\addplot[color=magenta,mark=*, only marks] coordinates {
	(1,2.4)
	(2,4.4)
	(3,6.4)
	(-1,1.6)
	(-2,3.6)
	(-3,5.6)
	};

\addplot[domain=0.2:4, color=cyan,thick]{2*x-0.4};

\addplot[domain=-4:0.2, color=cyan,thick, dashed]{-2*x+0.4};
\addplot[color=cyan,mark=*, only marks] coordinates {
	(-1,2.4)
	(-2,4.4)
	(-3,6.4)
	(1,1.6)
	(2,3.6)
	(3,5.6)
	};
\addplot[color=violet,mark=*, only marks] coordinates {
	(0,0.4)
	};

\end{axis}
\node[] at (4.5cm,6.8cm) {$H$};
\node[] at (9cm,1cm) {$\nu=p\tfrac{R_{\text{eff}}}{2\pi}$};

\node[] at (3.7cm,1.1cm) {$\color{magenta}Y(p)$};
\node[] at (4.8cm,1.1cm) {$\color{cyan}\bar{Y}(p)$};

\node[] at (6.2cm, 4cm) {$\color{cyan} J^{+}_{-\nu}$};

\node[] at (3.4cm, 4.6cm) {$\color{cyan} \tilde{J}^{-}_{-\tilde{\nu}}$};

\node[] at (2.2cm, 4cm) {$\color{magenta} \tilde{J}^{+}_{-\tilde{\nu}}$};

\node[] at (5.2cm, 4.6cm) {$\color{magenta} {J}^{-}_{-{\nu}}$};
\end{tikzpicture}
\caption{
Some excitations in the spectrally unflowed sector. We plot in red the dispersion $H(p)$ for the excitation $Y(p)$, and in cyan the one for $\bar{Y}(p)$. The position of the cusp where $kp=-\mu \Reff$ is close to zero---more precisely, it lies in the interval $(-1,+1)$. We highlight the allowed mode numbers with dots; the mode numbers falling on the positive slopes (solid lines) correspond to chiral excitations $J^\pm_{-\nu}$ while the ones on the negative slopes (dashed lines) correspond to anti-chiral ones, $\tilde{J}^\pm_{-\tilde{\nu}}$. The only allowed zero-modes are those of ${J}^-$ and $\tilde{J}^-$, as required by the WZW construction, see also appendix~\ref{app:WZW}.
}
\label{fig:unflowed}
\end{figure}

\paragraph{Identification of the mode~number.}
We have seen in eq.~\eqref{eq:Betheeqlinear} that the Bethe equations give a linear relation between momenta and excitation numbers $p_i \Reff=2\pi \nu_i$. 
Plugging these values into the dispersion~\eqref{eq:scdispersion}, we have
\begin{equation}
\label{eq:energyni}
H= \sum_i \left|\frac{k}{\Reff}\nu_i+\mu_i\right|\,.
\end{equation}
Which modes are chiral and which modes are anti-chiral depend on how we resolve the absolute value in this formula. Our na\"ive guess might be that positive modes correspond to chiral excitations, and negative modes to anti-chiral ones. More precisely, we should check whether
\begin{equation}
\nu_i>-\frac{\Reff}{k}\mu_i\quad\text{(chiral),}\qquad
\text{or}\qquad
\nu_i<-\frac{\Reff}{k}\mu_i\quad\text{(anti-chiral).}
\end{equation}
It is \emph{precisely because of the bound}~\eqref{eq:Rbound} that we can conclude that the modes are split as
\begin{equation}
\nu_i=1,2,\dots\quad\text{(chiral),}\qquad
\tilde{\nu}_i=-\nu_i=1,2,\dots\quad\text{(anti-chiral).}
\end{equation}
The same is also trivially true for the $\T^4$ bosonic excitations, which have $\mu=0$ and for which zero-modes should be discarded, see also the discussion in section~\ref{sec:sccomments}.
 We hence find that in the unflowed sector \textit{the spin-chain mode number and the WZW mode-number coincide}; more precisely
\begin{equation}
\NN=\Nsc\,,\qquad \NNR=\tilde{\Nsc}\,.
\end{equation}
Notice that a little more care is needed when analysing the zero-modes. Consider an excitation with $\mu_i=\pm1$ and $p_i=0$. In eq.~\eqref{eq:energyni} we should take the positive branch of the dispersion if $\mu_i=+1$ and the negative one if $\mu_i=-1$. Hence the zero-modes are split between chiral and anti-chiral representations in the WZW model. For instance, an $\su(2)$ excitation $Y(p)$ at $p=0$ corresponds to $J^-_0$ in the WZW model, see table~\ref{tab:bosonsmatch} for the complete matching of bosonic excitations. As the quantum numbers~$\delta \ell,\delta j$ and $N$ in table~\ref{tab:bosonsmatch} are additive, the bosonic spectrum in the unflowed sector of the WZW description has exactly the same form of the one of the spin chain when the bound~\eqref{eq:Rbound} is enforced.
Fermions require a little more work due to the usual subtleties in going from worldsheet to target-space supersymmetry. 

\renewcommand{\arraystretch}{1.5}
\begin{table}
\centering
\begin{tabular}{|c|c|c|c|c|c|}
\hline
$(s_\ell,s_j,s',s'')$ & Magnon & $m$ & $\su(2)_{\bullet}$ & $\su(2)_{\circ}$\\
\hline
\hline 
$(\, - \, - \, \mp \, \pm \, )$  & $ \ket{j_0 + \tfrac{1}{2}}^{\dot a}$ & 0 & $\boldsymbol 1$ & $\boldsymbol 2$ \\
\hline
$(\, - \, + \, \pm \, \pm \,)$  & $ \eta^a \chi_{\dot a} \ket{j_0 + \tfrac{1}{2}}^{\dot a}$ &  1  & $\boldsymbol 2$ & $\boldsymbol 1$ \\
\hline 
$(\, + \,- \, \pm \, \pm \,)$ &  $\epsilon_{\dot a \dot b } \eta^a \bar \chi^{\dot b} \ket{j_0 + \tfrac{1}{2}}^{\dot a} $ & 1 &  $\boldsymbol 2$ & $\boldsymbol 1$\\
\hline 
$(\, + \, + \, \mp \, \pm \,)$  & $\epsilon_{ab} \eta^a \eta^b \ket{j_0 + \tfrac{1}{2}}^{\dot a} $ & 2  & $\boldsymbol 1$ & $\boldsymbol 2$\\
\hline 
\end{tabular}
\caption{Level $\NN=0$ states in the chiral R sector of the WZW model and their spin-chain counterparts, which have momentum~$p=0$. In the WZW model they are identified by the labels $(s_\ell,s_j,s',s'')$, where $s_\ell$ keeps track of $\sl(2)$ fermionic zero-modes, $s_j$ of $\su(2)$ ones and $s',s''$ are related to the $\T^4$ directions. As discussed in appendix~\ref{app:WZW}, they take values $\pm1/2$ and $\delta=s_\ell+s_j+1$, which indeed matches the $m=\mu$ charge of the corresponding spin-chain excitation. We also indicate the relevant $\su(2)_\bullet\oplus\su(2)_\circ$ representations.}
\label{tab:fermion-matching-R}
\end{table}
\renewcommand{\arraystretch}{1}

\subsection{Fermionic excitations}

The identification of WZW (worldsheet) fermions with spin-chain excitations is reminiscent of the map between RNS fermions and GS states.  For the $\AdS{3}$ WZW models this was recently detailed in ref.~\cite{Dei:2018yth}. Loosely following that discussion, we will show that the identification of spin-chain and WZW charges $\delta = M$ extends to fermionic excitations. It is also possible to keep track of the $\so(4)_2=\su(2)_\bullet\oplus\su(2)_\circ$ charges, see also  appendix~\ref{app:WZW} for their description in the context of the WZW model. Before discussing the general argument in terms of partition functions, it is instructive to explicitly match some low-lying states in the chiral sector, namely those with~$\NN = 0$. Below we restrict to the unflowed chiral sector.

\paragraph{Chiral low-lying states ($\boldsymbol{\NN = 0}$).}

In the chiral sector at excitation number $\NN=0$ we have 8 fermionic states in the NS sector plus 8 in the R~sector, see appendix \ref{app:WZWBPS}.
As expected from our discussion of bosonic states, they can be mapped to spin-chain states containing only zero-momentum excitations. In particular, R-sector states are identified with spin-chain states containing  an even number of fermions while states in the NS sector contain an odd number of spin-chain excitations, see tables~\ref{tab:fermion-matching-R} and \ref{tab:fermion-matching-NS}, respectively.

\renewcommand{\arraystretch}{1.5}
\begin{table}
\centering
\begin{tabular}{|c|c|c|c|c|}
\hline
WZW state & Magnon & $M$ & $J_{\bullet}$ & $J_{\circ}$ \\
\hline 
\hline
$\psi^-_{-1/2} \ket{\ell_0, \ell_0 - 1}$ &  $\chi_{\dot a}  \ket{j_0 + \tfrac{1}{2}}^{\dot a}$  & 0 & $\boldsymbol 1$ & $\boldsymbol 1$ \\
\hline
$\theta^+_{-1/2} \ket{\ell_0, \ell_0 - 1}$ & $\epsilon_{\dot a \dot b} \bar \chi^{\dot b}  \ket{j_0 + \tfrac{1}{2}}^{\dot a}$ & 0  & $\boldsymbol 1$ & $\boldsymbol 1$ \\
\hline
$\gamma^{a \dot a}_{-1/2} \ket{\ell_0, \ell_0 - 1}$ & $\eta^a \ket{j_0 + \tfrac{1}{2}}^{\dot a}$  & 1 & $\boldsymbol 2$ & $\boldsymbol 2$ \\
\hline
$\theta^-_{-1/2} \ket{\ell_0, \ell_0 - 1}$ & $\epsilon_{ab} \eta^a \eta^b \chi_{\dot a} \ket{j_0 + \tfrac{1}{2}}^{\dot a}$  & 2 & $\boldsymbol 1$ & $\boldsymbol 1$ \\
\hline
$\psi^+_{-1/2} \ket{\ell_0, \ell_0 - 1}$ &  $ \epsilon_{ab} \epsilon_{\dot a \dot b} \eta^a \eta^b \bar \chi^{\dot b} \ket{j_0 + \tfrac{1}{2}}^{\dot a}$ & 2 & $\boldsymbol 1$ & $\boldsymbol 1$ \\
\hline 
\end{tabular}
\caption{Level $\NN=0$ fermions in the chiral NS sector of the WZW model and their spin-chain counterparts. As described in appendix~\ref{app:WZW}, the $\sl(2)$ fermions are denoted by~$\psi^\pm$, the $\su(2)$ ones by~$\theta^\pm$, and the four free fermions~$\gamma^{a\dot{a}}$ are bi-spinors of $\su(2)_\bullet\oplus\su(2)_\circ$.}
\label{tab:fermion-matching-NS}
\end{table}
\renewcommand{\arraystretch}{1}

\paragraph{Arbitrary (chiral) fermionic excitations.}

At higher level, an explicit construction such as the one presented above becomes quickly cumbersome. It is much more convenient to describe states with arbitrary mode number both in the WZW model and in the spin chain by writing down partition functions with suitable chemical potentials.
By virtue of the above discussion, it will be enough to match the states with a given mode number~$\NN$ and charge~$\delta$ to those in the spin-chain with mode number~$\Nsc$ and charge~$M$ to ensure the matching of the energy formula~\eqref{eq:WZW-unflowed-dispersion} with~\eqref{eq:SpinChainSpectrum}. We will actually be able to do more by individually matching all the $\sl(2)_\L$, $\su(2)_\L$ and $\so(4)_2=\su(2)_\bullet\oplus\su(2)_\circ$ charges of chiral excitations.
As with RNS and GS states, the identification can be performed by using Jacobi's ``abstruse identity'', see \textit{e.g.}\ chapter 21 of ref.~\cite{whittaker1996course}. This can be written as 
\begin{equation}
Z_{\text{GS}} = Z_{\text{NS}} + Z_{\text R}  \,,
\label{eq:abstruse-id}
\end{equation}
where the GS partition function features eight fermions
\begin{equation}
Z_{\text{GS}}  =  q^{\frac{1}{2}} \prod_{i=1}^4(z_i^{\frac{1}{2}} + z_i^{-\frac{1}{2}}) \, 
\prod_{n=1}^{\infty} (1 +  z_i \, q^n)\, (1 +  z_i^{-1} \, q^n) \,,
\end{equation}
distinguished by four chemical potential~$z_1,\dots z_4$. In the RNS description we have, instead%
\footnote{%
The difference of the two products in eq.~\eqref{eq:Z-NS} implements the NS-sector GSO projection by imposing that only states with an odd number of fermions appear. Similarly the sum in eq.~\eqref{eq:Z-R} imposes the GSO projection in the R sector, selecting states with even or odd fermion number according to the chosen configuration of zero modes, as explained in appendix~\ref{app:WZW}.  
}
\begin{equation}
Z_{\text {NS}}=\frac{1}{2}\Big[ \prod_{i=1}^4\prod_{n=1}^{\infty} (1 +  \zeta_i \, q^{n-\frac{1}{2}}) \, (1 +  \zeta_i^{-1}q^{n-\frac{1}{2}})
-
\prod_{i=1}^4\prod_{n=1}^{\infty} (1 -  \zeta_i \, q^{n-\frac{1}{2}}) \, (1 -  \zeta_i^{-1} q^{n-\frac{1}{2}})
\Big]
\label{eq:Z-NS}
\end{equation}
and
\begin{equation}
Z_{\text R}= \frac{q^{\frac{1}{2}}}{2}  \Big[ \prod_{i=1}^4 (\zeta_i^{- \frac{1}{2}} + \zeta_i^{ \frac{1}{2}}) \prod_{n=1}^{\infty} (1 +  \zeta_i \, q^n) \, (1 +  \zeta_i^{-1}q^n) \ + \  
\prod_{i=1}^4 (\zeta_i^{- \frac{1}{2}} - \zeta_i^{ \frac{1}{2}}) \prod_{n=1}^{\infty} (1 -  \zeta_i \, q^n) \, (1 -  \zeta_i^{-1}q^n) \Big] \,, 
\label{eq:Z-R}
\end{equation}
where the chemical potentials are labeled by $\zeta_1,\dots \zeta_4$. The abstruse identity dictates
\begin{equation}
z_i^2\,\zeta_i^2=\zeta_1\,\zeta_2\,\zeta_3\,\zeta_4\,,
\qquad \forall\ i=1,\dots 4\,.
\label{eq:abstruse-charges-relation}
\end{equation}
We can relate the chemical potential~$z_i$ to the  $\sl(2)_\L$, $\su(2)_\L$ and $\so(4)_2=\su(2)_\bullet\oplus\su(2)_\circ$ charges as it follows, see also table~\ref{tab:excitationszeromom}
\begin{equation}
\begin{aligned}
& \bar \eta^+ \sim z_1 = u_\ell^{-\frac{1}{2}} u_j^{\frac{1}{2}} u_\bullet^{\frac{1}{2}}  \,, \qquad &  &\eta^-  \sim z_1^{-1} = u_\ell^{\frac{1}{2}} u_j^{-\frac{1}{2}} u_\bullet^{-\frac{1}{2}}   \,, \\
& \eta^+ \sim z_2 = u_\ell^{\frac{1}{2}} u_j^{-\frac{1}{2}} u_\bullet^{\frac{1}{2}}  \,, \qquad &  & \bar \eta^-  \sim z_2^{-1} = u_\ell^{-\frac{1}{2}} u_j^{\frac{1}{2}} u_\bullet^{-\frac{1}{2}}   \,, \\
& \bar \chi^+ \sim z_3 = u_\ell^{\frac{1}{2}} u_j^{\frac{1}{2}} u_\circ^{\frac{1}{2}}  \,, \qquad &  & \chi^-  \sim z_3^{-1} = u_\ell^{-\frac{1}{2}} u_j^{-\frac{1}{2}} u_\circ^{-\frac{1}{2}}   \,,\\
& \bar \chi^- \sim z_4 = u_\ell^{\frac{1}{2}} u_j^{\frac{1}{2}} u_\circ^{-\frac{1}{2}}  \,, \qquad &  & \chi^+  \sim z_4^{-1} = u_\ell^{-\frac{1}{2}} u_j^{-\frac{1}{2}} u_\circ^{\frac{1}{2}}   \,.
\end{aligned}
\end{equation}
Similarly, from appendix~\ref{app:WZW} and from the caption of table~\ref{tab:fermion-matching-R} we can read off the chemical potentials and the charges of RNS fermions. Indeed we find
\begin{equation}
\begin{aligned}
& \psi^{\pm}\sim \zeta_1^{\pm1} = u_\ell^{\pm1} \,,
& & \gamma^{\pm \mp}\sim\zeta_3^{\pm1}=u_\bullet^{\pm1/2}u_\circ^{\mp1/2}
 \,, \\
& \theta^{\pm}\sim \zeta_2^{\pm1} = u_j^{\pm1} \,, \qquad 
& &  \gamma^{\pm\pm}\sim\zeta_4^{\pm1}=u_\bullet^{\pm1/2}u_\circ^{\pm1/2}
 \,.
\end{aligned}
\end{equation}
It is immediate to verify that these chemical potentials satisfy eq.~\eqref{eq:abstruse-charges-relation} so that the spin-chain and WZW partition functions match by virtue of Jacobi's abstruse identity.
As for the charges of the fermionic zero-modes, which in eq.~\eqref{eq:Z-R} as are encoded into the terms $ \zeta_i^{- \frac{1}{2}} \pm \zeta_i^{ \frac{1}{2}}$, we get a product of the form%
\footnote{Notice that, as usual, the identification of the $\su(2)$ quantum numbers involves an overall sign.}
\begin{equation}
\zeta_1^{s_\ell}\,
\zeta_2^{-s_j}\,
\zeta_3^{s'}\,
\zeta_4^{s''} =
u_\ell^{s_\ell}\,u_j^{-s_j}
\,u_{\bullet}^{(s'+s'')/2}\,u_{\circ}^{(s''-s')/2}\,,
\end{equation}
consistently with eq.~\eqref{eq:sjrelation}.

\subsection{Spectrally-flowed sectors}

More general representations of the WZW model can be constructed by spectral flow as reviewed in appendix~\ref{sec:flowed-sector}. It is convenient to  perform the spectral flow in the same way in the $\sl(2)_k$ and in the $\su(2)_k$ algebras. This ``supersymmetric'' way of flowing will make it particularly easy to identify the BPS spectrum, as discussed in the appendix. Moreover, remark that modular invariance requires us to take the same spectrally flowed representations (and hence in this case the same flow parameter~$w$) in the chiral and anti-chiral sector. Then we find that the light-cone energy is given by
\begin{equation}
\label{eq:mass-shell-flowed-symm}
\Htot=\sqrt{(2j_0+k w+1)^2 + 4 k (\Neff- w \delta)}-(2j_0+k w+1)+\delta +\tilde{\delta},
\end{equation}
where the spectral-flow parameter is $w\in\mathbb{N}$ and $\delta=\delta\ell +\delta j+s_\ell+s_j+1$, like in the unflowed sector---and similarly for~$\tilde{\delta}$. Below we shall see how this formula emerges from the spin-chain description.

\paragraph{Spin-chain vacuum and spin-chain length.} Building on the previous subsection, it is natural to identify
\begin{equation}
R_0=2j_0+1+kw\,.
\end{equation}
This too gives a BPS state; it is the image of the R-R BPS state considered above in the $w$-th spectrally-flowed sector. Due to the Maldacena-Ooguri bound, and following the discussion above, we find that the effective spin-chain length is bounded by
\begin{equation}
\label{eq:flowed-bound}
\frac{1}{2}+kw<\Reff<k(w+1)-\frac{1}{2}\,.
\end{equation}
Given that we have performed the spectral flow in the same way in  $\sl(2)_k$ and in $\su(2)_k$, the discussion of the BPS states follows the one above.

\begin{figure}
\centering
\begin{tikzpicture}
\begin{axis}[xlabel={},ylabel={},
height=8cm, width=14cm,
xmin=-7,xmax=7,ymin=-1,ymax=7, axis lines=center, yticklabels={,,}, axis equal]

\addplot[domain=-2.7:4, color=magenta,thick]{2*x+5.4};

\addplot[domain=-7:-2.7, color=magenta,thick, dashed]{-2*x-5.4};

\addplot[color=magenta,mark=*, only marks] coordinates {
	(-3,0.6)
	(-4,2.6)
	(-5,4.6)
	(-6,6.6)
	(-2,1.4)
	(-1,3.4)
	};

\addplot[domain=2.7:7, color=cyan,thick]{2*x-5.4};

\addplot[domain=-1:2.7, color=cyan,thick, dashed]{-2*x+5.4};
\addplot[color=cyan,mark=*, only marks] coordinates {
	(3,0.6)
	(4,2.6)
	(5,4.6)
	(6,6.6)
	(2,1.4)
	(1,3.4)
	};
\addplot[color=violet,mark=*, only marks] coordinates {
	(0,5.4)
	};

\end{axis}
\node[] at (6.5cm,6.8cm) {$H$};
\node[] at (13cm,1cm) {$\nu=p\tfrac{R_{\text{eff}}}{2\pi}$};

\node[] at (3.85cm,0.4cm) {$\color{magenta}Y(p)$};
\node[] at (8.5cm,0.4cm) {$\color{cyan}\bar{Y}(p)$};

\node[] at (10.5cm, 4cm) {$\color{cyan} J^{+}_{-\nu}$};

\node[] at (7.5cm, 3.6cm) {$\color{cyan} \tilde{J}^{-}_{-\tilde{\nu}}$};

\node[] at (2cm, 4cm) {$\color{magenta} \tilde{J}^{+}_{-\tilde{\nu}}$};

\node[] at (4.8cm, 3.6cm) {$\color{magenta} {J}^{-}_{-{\nu}}$};
\end{tikzpicture}
\caption{
Some excitations in the flowed sector with $w=2$. We plot in red the dispersion $H(p)$ for the excitation $Y(p)$, and in cyan the one for $\bar{Y}(p)$. The position of the cusp where $kp=-\mu \Reff$ is between $-3$ and $-2$, or between $2$ and $3$. We highlight the allowed mode numbers with dots; the mode numbers falling on the positive slopes (solid lines) correspond to chiral excitations $J^\pm_{-\nu}$ while the ones on the negative slopes (dashed lines) correspond to anti-chiral ones, $\tilde{J}^\pm_{-\tilde{\nu}}$. Notice that the mode-number is shifted with respect to figure~\ref{fig:flowedspectrum}.
}
\label{fig:flowedspectrum}
\end{figure}
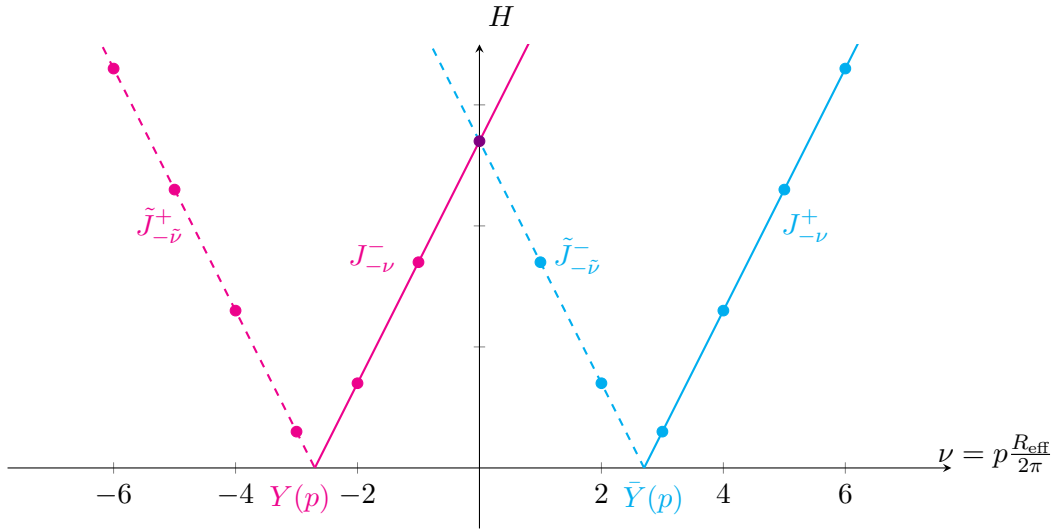

\paragraph{Identification of the mode~number.}
Much like above, the relation between momentum and spin-chain mode-number is a linear one, $p_i = 2\pi \nu_i/\Reff$. A subtlety arises when identifying chiral and anti-chiral excitations from eq.~\eqref{eq:energyni}. The condition on the mode-numbers is again
\begin{equation}
\nu_i>-\frac{\Reff}{k}\mu_i\quad\text{(chiral),}\qquad
\text{or}\qquad
\nu_i<-\frac{\Reff}{k}\mu_i\quad\text{(anti-chiral).}
\end{equation}
Given that $w\lesssim{\Reff}/{k}\lesssim w+1$
by virtue of the bound~\eqref{eq:flowed-bound}, we have that, for $\mu_i=+1$
\begin{equation}
\nu_i=-w, -w+1, \dots\,\quad\text{(chiral)},
\qquad
\tilde{\nu}_i=-\nu_i=w+1, w+2, \dots\,\quad\text{(anti-chiral)},
\end{equation}
whereas for $\mu_i=-1$ we have
\begin{equation}
\nu_i=w+1, w+2, \dots\,\quad\text{(chiral)},
\qquad
\tilde{\nu}_i=-\nu_i=-w, -w+1, \dots\,\quad\text{(anti-chiral)}.
\end{equation}
In other words, the notion of chiral and anti-chiral mode number is shifted by $w\,\mu_i$, as illustrated in figure~\ref{fig:flowedspectrum}.
For example, let us consider an $\su(2)$ excitation $Y(p)$ with the lowest allowed \textit{chiral} mode number, so that $p=-2\pi w/\Reff$. This corresponds to acting with the lowest-moded $J^-$ oscillator in the flowed representation, namely $J^{-}_{w}$; higher modes are $J^-_{w-1}, J^-_{w-2}$ and so on. For the \textit{anti-chiral} mode instead we look at $\bar{Y}(p)$ the highest allowed momentum in the anti-chiral sector, that is $p=2\pi\,w/\Reff$. This corresponds to acting with $\tilde{J}^{-}_{w}$. 
Conversely, for $\bar{Y}(p)$, the lowest \textit{chiral} mode number gives~$p=-2\pi\,w/\Reff$. This corresponds to acting with the lowest $J^+$ mode in the flowed representation, that is $J^{+}_{-w-1}$. This precisely matches the structure of the spectrally flowed $\su(2)_k$ representation, which is summarised in eq.~\eqref{eq:flowed-rep-su2} for $\su(2)_k$; similar considerations apply for $\sl(2)_k$ excitations. We describe the map between general bosonic excitations in table~\ref{tab:bosonsmatchflowed}. The discussion for fermions is similar to the one of the unflowed sector and we omit it.
In summary, the spin-chain mode number is shifted with respect to the WZW one, in a way dependent on the charge $\mu_i$,
\begin{equation}
\nu_i = n_i-w\,\mu_i\,,
\end{equation}
and similarly in the anti-chiral sector, so that
\begin{equation}
\label{eq:simpleflowN}
\Nsc=\NN - w\,\delta\,,
\qquad
\tilde{\Nsc}=\NNR-w\,\tilde{\delta}\,.
\end{equation}
Notice that the level-matching condition, which in this sector is simply $\Ptot=0$ in terms of the WZW excitation numbers reads
\begin{equation}
\NN - w\,\delta=\NNR-w\,\tilde{\delta}\,,
\end{equation}
which is exactly eq.~\eqref{eq:flowed-level-matching}. Finally, in terms of these variables the WZW mass-shell condition~\eqref{eq:mass-shell-flowed-symm} becomes
\begin{equation}
\Htot=\sqrt{R_0^2 + 4 k\, \Nsc}-R_0+\delta +\tilde{\delta},
\end{equation}
in perfect agreement with the spin-chain formula~\eqref{eq:SpinChainSpectrum-W0}, bearing in mind that $\delta=M$ and $\tilde{\delta}=\tilde{M}$, which follows from the same argument as in the previous section.

\renewcommand{\arraystretch}{1.2}
\begin{table}[t]
\centering
\begin{tabular}{|l|c|c|c|c|l|c|}
\hline
WZW & $\delta \ell$ & $\delta \tilde \ell$ & $\delta j$ & $\delta \tilde \jmath$ & Spin chain & $m+\tilde{m}$ \\
\hline
$L_{-n+w+1}^+$ & $+ 1$ &  &  &  & $Z \big(2 \pi \frac{n-w-1}{\Reff}\big)$ & $+ 1$ \\
$L_{-n-w}^-$ & $- 1$ &  &  &  & $\bar Z \big(2 \pi \frac{n+w}{\Reff}\big)$ & $- 1$ \\
$J_{-n-w}^+$ &  & & $- 1$ &  & $\bar Y \big(2 \pi \frac{n+w}{\Reff}\big)$ & $- 1$ \\
$J_{-n+w+1}^-$ &  & & $+ 1$ &  & $ Y \big(2 \pi \frac{n-w-1}{\Reff}\big)$ & +1 \\
\hline
$\tilde L_{-n + w + 1}^+$ &  & $+ 1$ &  &  & $\bar Z \big(-2 \pi \frac{n-w-1}{\Reff}\big)$ & $+ 1$  \\
$\tilde L_{-n-w}^-$ &  & $- 1$ &  &  & $ Z \big(-2 \pi \frac{n+w}{\Reff}\big)$ & $- 1$  \\
$\tilde J_{-n-w}^+$ &  &  &  & $- 1$ & $Y \big(-2 \pi \frac{n+w}{\Reff}\big)$ & $- 1$  \\
$\tilde J_{-n+w+1}^-$ &  &  &  & $+ 1$ & $\bar Y \big(-2 \pi \frac{n-w-1}{\Reff}\big)$ & $+1$  \\
\hline
\end{tabular}
\caption{The spin-chain bosonic excitations are matched quite straightforwardly with those in the $w$-th flowed sector of the WZW model.
As explained in the text and illustrated in figure~\ref{fig:flowedspectrum}, it is necessary to shift the mode numbers by $\pm w$ depending on the value of~$\mu$ or, in the WZW language, on the charge of the Ka\v{c}-Moody current. As a result, it is possible to identify the shifted modes, with $n\geq1$.}
\label{tab:bosonsmatchflowed}
\end{table}
\renewcommand{\arraystretch}{1} 

\subsection{(Light-cone) winding sectors}
It remains to describe the sectors that in eq.~\eqref{eq:SpinChainSpectrum} have $W\neq0$, and hence non-vanishing total momentum
\begin{equation}
\label{eq:windingsc}
\Ptot = \frac{2\pi}{R_0}(\Nsc-\tilde{\Nsc})=2\pi W\,.
\end{equation}
In view of the discussion in appendix~\ref{app:lcgauge}, see in particular eq.~\eqref{eq:windinglc}, it is natural to associate such sectors with those that in light-cone gauge have non-trivial winding along the compact light-cone direction.

In order to identify such states in the WZW description we can build on the intuition developed in the previous section where we considered the spectrally flowed sectors. We have seen that the relation between $\Neff$ in the WZW model and $\Nsc$ undergoes a shift~\eqref{eq:simpleflowN}, which accounts for the mode-shift described in figure~\ref{fig:flowedspectrum}.
Given that in the WZW model $\Neff=\tilde{N}_{\text{eff}}$, in order to reproduce eq.~\eqref{eq:windingsc} we should perform the spectral flow differently in the chiral and anti-chiral algebras. As we review in appendix~\ref{sec:flowed-sector}, this is not possible for~$\sl(2)_k$, but it does make sense for $\su(2)_k$ as long as we require that the difference of the two $\su(2)_k$ spectral flow parameters $w_L-w_R$ is even. Moreover, we will require that the average $w=\tfrac{1}{2}(w_L+w_R)$, which is also an integer, is taken as spectral flow parameter of $\sl(2)_k$.  This is as close as we can get to the ``supersymmetric'' spectral flow that we have used in the section above. As derived in the appendix, see eq.~\eqref{eq:l0-asymmetric} and below, we find the light-cone energy
\begin{equation}
\begin{aligned}
\Htot&=&\sqrt{ \Big(2j_0 + 1 + \frac{w_L + w_R}{2}k \Big)^2 + 2 k (\mathcal{N} +   \tilde{\mathcal{N}}) +  \Big(\frac{w_L - w_R}{2}\Big)^2k^2 }
\qquad\qquad\qquad\\
&&-
\Big(2j_0 + 1 + \frac{w_L + w_R}{2}k \Big)+\delta+\tilde{\delta}\,,
\end{aligned}
\end{equation}
which conveniently is already written in terms of the spin-chain excitation numbers. Those are related to the WZW ones by
\begin{equation}
\label{eq:finalformula}
\begin{aligned}
\mathcal{N} = \NN - w \big(\delta\ell+s_\ell +\frac{1}{2}\big)  - w_L \big(\delta j+s_j +\frac{1}{2}\big)\,, \\
\tilde{\mathcal{N}} = \NNR - w \big(\delta\tilde{\ell}+\tilde{s}_\ell +\frac{1}{2}\big)  - w_R \big(\delta \tilde{\jmath}+\tilde{s}_j +\frac{1}{2}\big)\,,
\end{aligned}
\end{equation}
where as usual with a slight abuse of notation we simply set the R-sector spins $s_\ell$, $s_j$ to zero in the NS sector. As mentioned in the appendix, $\Nsc$ and $\tilde{\Nsc}$ are integer owing to the GSO projection~\cite{Ferreira:2017pgt}.
This matches the spin-chain energy~\eqref{eq:SpinChainSpectrum} on the nose with
\begin{equation}
\label{eq:finalmatch}
R_0 = 2 j_0 +1 + \dfrac{w_R+w_L}{2} k\,,\qquad
W=\frac{w_L-w_R}{2}\in\mathbb{Z}\,.
\end{equation}
As expected $R_0$ is the total R~charge of the ``vacuum'' after spectral flow. It is worth noting that, much like in the Green-Schwarz formalism, in presence of non-zero winding the would-be vacuum state with $\delta=\Nsc=0$ and $\tilde{\delta}=\tilde{\Nsc}=0$ is not BPS as evident from eq.~\eqref{eq:finalformula}. In fact, it is not even a physical state. In the spin-chain it is obvious that we cannot have a vacuum in presence of macroscopic total momentum $\Ptot\neq0$, as this is supplied by (typically many) excitations. The same is true in the WZW model: the level-matching constraint
\begin{equation}
\mathcal{N} - \tilde{\mathcal{N}} = \dfrac{w_R-w_L}{2} \left( 2 j_0 +1 +  \dfrac{w_R+w_L}{2} k \right)
\end{equation}
ensures that we cannot take $\Nsc=\tilde{\Nsc}=0$ when $\tfrac{1}{2}(w_L+w_R)\neq0$. What is more, using the identification~\eqref{eq:finalmatch} we find that the level-matching constraint reduces to
\begin{equation}
\mathcal{N} - \tilde{\mathcal{N}} = W\,R_0\,.
\end{equation}
which is precisely the spin-chain quantisation condition for the total momentum~\eqref{eq:SpinChainSpectrum}.

\section{Conclusions}
\label{sec:conclusions}
Building on ref.~\cite{Baggio:2018gct}, we have constructed an integrable spin-chain which describes strings on $\AdSST$ pure-NS-NS backgrounds. We have shown that the quantisation conditions on the energy spectrum which follow from the Bethe ansatz match the physical-state conditions of the WZW model on the nose.
It would be very interesting to further explore this correspondence in order to shed more light on the relation between integrability and $\CFT_2$ techniques. One obvious question is how the underlying algebraic structure of the two models is related. For the spin~chain, this is given by the algebraic Bethe ansatz operators (or, in the GS worldsheet picture, by the Zamolodchikov-Faddeev algebra); for the WZW model it is given by the Ka\v{c}-Moody algebra. Establishing a direct relation between these two structures would be extremely interesting. It would also be interesting to see in more detail how the unitarity and null-state conditions of the Ka\v{c}-Moody representations emerge from the spin~chain.

It might also be instructive to consider the semi-classical limit of this spin~chain.  One obvious class of relevant classical solutions is given by giant magnons~\cite{Hofman:2006xt,Hoare:2013lja}, though many other classical solutions have been investigated in presence of NS-NS flux, see \textit{e.g.} refs.~\cite{David:2014qta,Banerjee:2014gga, Banerjee:2015bia,Banerjee:2015qeq}. Furthermore, it is known that certain spinning string solutions can be described by the Neumann-Rosochatius integrable system~\cite{Arutyunov:2003uj,Arutyunov:2003za}. Indeed this has been investigated in some detail in the context of mixed-flux $\AdSST$ backgrounds~\cite{Hernandez:2014eta,Hernandez:2015nba, Hernandez:2018gcd,Nieto:2018jzi}. Furthermore, it was found that the integrability description simplifies in the pure-NS-NS limit~\cite{Hernandez:2014eta,Nieto:2018jzi}, like it happens here. It is therefore natural to wonder how the Neumann-Rosochatius model is related to this spin~chain.

We expect the correspondence between integrability and stringy WZW models to go beyond the spectrum of the theory and to include three- and higher-point correlation functions of protected and non-protected states. A recently-developed integrability technique, the ``hexagon form factor'' approach~\cite{Basso:2015zoa}, allows in principle for the computation of closed-string three-point functions for integrable backgrounds. Moreover, the technique can be extended to higher-point functions~\cite{Eden:2016xvg,Fleury:2016ykk} and in principle to non-planar observables~\cite{Eden:2017ozn,Bargheer:2017nne}. While it is not obvious whether the hexagon form factor can be derived in the present context, it is encouraging to notice that the wrapping corrections, whose description is a major obstacle for hexagon program~\cite{Basso:2015zoa,Eden:2015ija,Basso:2015eqa,Basso:2017muf} here seem to be almost inconsequential, at least for two-point functions. Indeed for the $\AdSST$ there exist closed formulae for correlation functions~\cite{Teschner:1997ft,Teschner:1999ug, Maldacena:2001km, Giribet:2007wp, Cardona:2009hk,Cardona:2010qf}---a far cry from the $\AdS{5}/\CFT_4$ case! Therefore, this looks like an ideal playground to test these techniques.

On top of this, it would be interesting to build on the spin~chain discussed here to try to better understand the integrability properties of the \textit{dual} $\CFT_2$, that is the symmetric-product orbifold CFT of~$\T^4$, $\text{Sym}_N\T^4$~\cite{Maldacena:1997re}. A study of the symmetries of that CFT suggests that it is related to a spin-chain of the type considered here with parameter $k=1$~\cite{Sfondrini:talk}. Indeed very recently it was argued that the spectrum of the level~$k=1$ WZW model contains that of the symmetric-product orbifold~\cite{Giribet:2018ada, Gaberdiel:2018rqv}. It would hence be quite interesting to explore this spin~chain in the small-$k$~regime.

It also seems very plausible that this integrability approach can be extended to more general pure-NS-NS $\AdS{3}$ backgrounds, such as $\AdS{3}\times\S^3\times\text{K3}$ and $\AdS{3}\times\S^3\times\S^3\times\S^1$ as well as to flat-space strings and superstrings, whose integrability properties have been to some extent investigated recently~\cite{Dubovsky:2012wk, Cooper:2014noa,Mohsen:2016lch}. In general we expect that for these theories the S~matrix will take the same form, but the spectrum of fundamental excitations will be different---\textit{e.g.}, the masses~$\mu$ will take more general values~\cite{Dei:2018jyj}, see also refs.~\cite{Dubovsky:2012wk,Borsato:2012ud, Borsato:2012ss,Borsato:2015mma}; moreover, non-trivial integrable boundary conditions may be imposed for the string NLSM fields, resulting in orbifolds.
Similarly, integrable deformations of these backgrounds, such as TsT~\cite{Frolov:2005dj} and quantum deformations~\cite{Delduc:2013qra}, see \textit{e.g.}\ refs.~\cite{vanTongeren:2013gva,Borsato:2016hud} for reviews, could be studied.
Finally, as the simplicity of the theory we investigated here can be seen~\cite{Baggio:2018gct} as stemming from its relation to the (generalised) $T\bar{T}$ deformation~\cite{Zamolodchikov:2004ce,Dubovsky:2013ira, Smirnov:2016lqw,Cavaglia:2016oda} of a \textit{free} theory, it is natural to ask whether any more general such deformation might be physically relevant, see also refs.~\cite{Bonelli:2018kik, Conti:2018jho} for recent work in this direction.
We hope to return soon to some of these questions.

\section*{Acknowledgements}
We thank Marco Baggio, Andrea Cavagli\`a, Sergei Dubovsky, Marius de Leeuw, Matthias Gaberdiel and Sergey Frolov for useful related discussions. We are especially grateful to Marco Baggio, Andrea Cavagli\`a, Matthias Gaberdiel and Roberto Tateo for their comments on a preliminary version of this article.   This work is partially supported through a research grant of the Swiss National Science Foundation, as well as by the NCCR SwissMAP, funded by the Swiss National Science Foundation. A.S.\ also acknowledges support by the ETH ``Career Seed Grant'' no.~0-20313-17.

\newpage
\appendix
\section{Uniform light-cone gauge}
\label{app:lcgauge}
Let us briefly review how uniform light-cone gauge~\cite{Arutyunov:2004yx,Arutyunov:2005hd,Arutyunov:2006gs} is fixed for $\AdSST$, following and slightly generalising refs.~\cite{Baggio:2018gct,Borsato:2014hja,Lloyd:2014bsa}. For definiteness, let us normalise the bosonic part of the string NLSM action to
\begin{equation}
\mathcal{S}=-\frac{k}{4\pi}\int\limits_0^\infty\de\tau \int\limits_0^R \de\sigma\, (\gamma^{\alpha\beta}G_{\mu\nu} +\epsilon^{\alpha\beta}B_{\mu\nu})\,\partial_\alpha X^\mu\partial_\beta X^\nu\,,
\end{equation}
where $G_{\mu\nu}$ is the $\AdSST$ metric with unit-radius $\S^3$, $B_{\mu\nu}$ is the Kalb-Ramond field in the normalisation of ref.~\cite{Baggio:2018gct}, and $\gamma^{\alpha\beta}$ is unit-determinant worldsheet metric.
Let~$t$ be the isometric time coordinate in $\AdS{3}$ and $\phi$ be a great circle on~$\S^3$. For $0\leq a\leq 1$ we introduce the light-cone coordinates
\begin{equation}
x^+=(1-a)\,t+a\,\phi\,,\qquad
x^-=\phi-t\,.
\end{equation}
In this notation we fix light-cone gauge as it follows:
\begin{equation}
x^+=\tau +\frac{2\pi\,a}{R} W,
\qquad
p_-=(1-a)p_\phi-a p_t=1\,,
\end{equation}
where we used the conjugate momenta $p_\mu=\delta\mathcal{S}/\delta (\partial_0 X^\mu)$.
This gauge choice slightly generalises the one of ref.~\cite{Baggio:2018gct} in that it allows for winding in the light-cone direction; more precisely
\begin{equation}
\phi(\sigma)-\phi(\sigma+R)=2\pi W\,,\qquad
W\in\mathbb{Z}\,.
\end{equation}
where~$R$ is the size of the worldsheet, see ref.~\cite{Arutyunov:2004yx}.

In this gauge-fixing, the classical light-cone Hamiltonian is
\begin{equation}
H_{\text{cl.}}=-\int\limits_0^R\de\sigma\, p_+=-\int\limits_0^R\de\sigma \,(p_t+p_\phi)\,,
\end{equation}
for any $a$, which gives eq.~\eqref{eq:lcHam}. On the other hand the length of the worldsheet is fixed by
\begin{equation}
R=\int\limits_0^R\de\sigma\, p_-
= (1-a)\int\limits_0^R\de\sigma\, p_\phi+a\int\limits_0^R\de\sigma \,p_t\,.
\end{equation}
In the $a=0$ gauge, which as discussed in ref.~\cite{Baggio:2018gct} is where the worldsheet S~matrix is simplest, the length is the total R~charge
\begin{equation}
R=j+\tilde{\jmath}\,.
\end{equation}
Finally, the level-matching constraint arises from imposing appropriate boundary conditions for the transverse field $x^-$. Using the fact that in light-cone gauge this coordinate is related to the momentum density on the worldsheet,  $\partial_\sigma x^- = -p_{\text{w.s.}}$, we have
\begin{equation}
\label{eq:windinglc}
2\pi W = -\int\limits_0^R\de\sigma\,\partial_\sigma x^-= \Ptot\,.
\end{equation}

\section{Different ``frames'' for the worldsheet S~matrix}
\label{app:Smatframe}
The worldsheet S~matrix is not an observable. Rather, it depends on the choice of gauge (see appendix~\ref{app:lcgauge}) and on the \textit{frame}.
A perturbative worldsheet computation such as the one of ref.~\cite{Hoare:2013pma} is performed in the so-called string frame~\cite{Arutyunov:2006yd}. This means that in a suitable ($a=0$) gauge the length of the worldsheet is given by the R~charge of a given state, $R=j+\tilde{\jmath}$, and the S~matrix takes the form
\begin{equation}
\label{eq:Smatrix-stringframe}
\Smat_{\text{string}}(p_1,p_2)=e^{\frac{i}{2}\Phi(p_1,_2)}\, \Umat(p_2)\otimes \Umat^\dagger(p_1)\,,
\end{equation}
where $\Phi(p_1,p_2)$ is given by eq.~\eqref{eq:cddfactor}. Notice that this expression differs from eq.~\eqref{eq:Smatrix} due to the presence of the matrix $\Umat$, which acts on the vector-space of the first and second particle separately. Notice also that the action on the vector-space of the first particle depends on the momentum of the second particle, and viceversa. The matrix $\Umat$ itself is diagonal and given in terms of the R~charge of the state on which it acts, namely
\begin{equation}
\label{eq:Umat}
\Umat(p) = \exp\big[{i p\,(\Jgen^3+\tilde{\Jgen}^3)}\big]\,.
\end{equation}
Hence in the string frame, the S~matrix is diagonal, but not proportional to the identity; in components,
\begin{equation}
\Big[\Smat_\text{string}(p_i,p_j)\Big]_{ij}^{kl}= S^\text{string}_{ij}(p_i,p_j)\, \delta_{i}^k\,\delta_j^l\,.
\end{equation}

Matrices of the form~\eqref{eq:Umat} can be interpreted as twists of the Zamolodchikov-Faddeev (ZF) algebra~\cite{Arutyunov:2006yd}. More specifically, given an S~matrix satisfying the ZF algebra one can introduce a \textit{twisted} S-matrix by the conjugation
\begin{equation}
\label{eq:ZFtwist}
\Smat_\text{string}(p_1,p_2)\to
\Smat_\text{twist}(p_1,p_2) =\Umat^\dagger(p_2)\otimes \mathbf{1}\cdot \Smat_\text{string}(p_1,p_2) \cdot \mathbf{1}\otimes \Umat(p_1)\,.
\end{equation}
Such a twist cannot be induced by a change of basis on the space of the $(8|8)$ fundamental modes; rather, it comes from a redefinition of the \textit{two-particle} module, and is tantamount to a twist in the co-product on the underlying algebra, see \textit{e.g.}~\cite{Arutyunov:2009ga} for a review.
In our particular case, it is clear that there exists a twist of the ZF algebra such that the S~matrix is not only diagonal, but \textit{proportional to the identity}. This can be done by acting as in eq.~\eqref{eq:ZFtwist} precisely with the matrix of eq.~\eqref{eq:Umat}. Moreover, in the context of $\AdS{3}/\CFT_2$, matrices of the form~\eqref{eq:Umat} are associated to a \textit{frame change} from the ``string frame'' to the ``spin-chain'' frame.

The meaning of these frames becomes clearer if we write down the Bethe-Yang equations for a state with $K$ excitations: in the sector of zero light-cone winding, \textit{cf.}\ eq.~\eqref{eq:windinglc}, we have
\begin{equation}
1=e^{i p_i(j+\tilde{\jmath})}\prod_{j\neq i}^K S_{ij}^{\text{string}}(p_i,p_j)\,,\qquad i\dots K;\qquad \sum_{i=1}^K  p_i=0\,.
\end{equation}
More generally, in presence of non-trivial winding we have 
\begin{equation}
\label{eq:BetheYang-string-frame}
(-1)^{W\,F_i}=e^{i p_i(j+\tilde{\jmath})}\prod_{j\neq i}^K S_{ij}^{\text{string}}(p_i,p_j)\,,\qquad i\dots K;\qquad \sum_{i=1}^K  p_i=2\pi W\,,
\end{equation}
where $F_i$ is the fermion number of the $i$-th excitation; this accounts for the fact that fermions are anti-periodic in odd-winding sectors, see \textit{e.g.}\ ref.~\cite{Arutyunov:2009ga}.
Let us now introduce the labels $(\delta j_i,\delta \tilde{\jmath}_i)$ for the (left and right) R~charge of the $i$-th excitation, so that
\begin{equation}
R=j+\tilde{\jmath} = R_0+\sum_{i=1}^K (\delta j_i+\delta \tilde{\jmath}_i)\,,
\end{equation}
where $R_0$ is the R~charge \textit{of the BPS vacuum}. Using these expression we can write explicitly the action of $\Umat(p)$ on a single excitation as
\begin{equation}
\Big[\Umat(p)\Big]_k^l=e^{ip \,\delta j_k}\delta_k^l\,,
\end{equation}
and explicitly rewrite the Bethe-Yang equations as
\begin{equation}
\begin{aligned}
(-1)^{W\,F_i}&=&&e^{i p_i\, R}\prod_{k=1}^K e^{-i p_i (\delta j_k+\delta \tilde{\jmath}_k)+i p_k (\delta j_i+\delta \tilde{\jmath}_i)} e^{\frac{i}{2}\Phi(p_i,p_k)}\\
&=&&e^{i p_i \,R_0}e^{i (\delta j_i +\delta \tilde{\jmath}_i)\Ptot}\prod_{k=1}^K e^{\frac{i}{2}\Phi(p_i,p_k)}=
e^{i p_i \,R_0}e^{i 2\pi(\delta j_i +\delta \tilde{\jmath}_i)W}\prod_{k=1}^K e^{\frac{i}{2}\Phi(p_i,p_k)}\,.
\end{aligned}
\end{equation}
Observing that the total R~charge is half-integer precisely for fermions (see also table~\ref{tab:excitationszeromom}), we recover precisely the Bethe equations~\eqref{eq:betheeqs}.
In these Bethe equations, equivalent to~\eqref{eq:BetheYang-string-frame}, the scattering is given by the \textit{twisted S~matrix} $S_\text{twist}=e^{i\Phi/2}$ only, and the string length is $R_0$, rather than $R=j+\tilde{\jmath}$.
The conclusion is that the twist~\eqref{eq:ZFtwist} takes us from the \textit{string frame} where system's size is \textit{the R-charge of each state} $j+\tilde{\jmath}$ (as natural in light-cone gauge, see appendix~\ref{app:lcgauge}), to the spin-chain frame where the volume of the integrable system is the R~charge \textit{of the vacuum}~$R_0$.
The fact that in the spin-chain frame the S~matrix is substantially simpler than in the string one is another hint that a spin-chain interpretation is particularly natural for this integrable system.

\section{Mirror TBA equations}
\label{app:mTBA}
Given an integrable theory at finite temperature $1/R$, its free energy can be computed by the TBA~\cite{Yang:1968rm}. Let us briefly review how to do this for a non-relativistic theory of bosons and fermions with diagonal scattering, see also ref.~\cite{Bajnok:2010ke}.
Since we are in large volume $\beta\to\infty$, the mirror Bethe-Yang equations are almost exact. They impose that the mirror momentum~$\bar{p}$ of a particle of flavour $a$ satisfies%
\footnote{%
Here we are assuming that there exist no non-trivial ``Bethe strings'', which is quite natural given the linear form of the dispersion. See \textit{e.g.}~\cite{Arutyunov:2009zu,Bombardelli:2009ns} for a discussion of this point for $\AdS{5}\times\S^5$ strings.
}
\begin{equation}
\bar{p}_{a}\, \beta + \sum_{j=1}^M \varphi_{a\,b_j}(\bar{p}_{a},\bar{p}_{j,b_j})= 2\pi n_{a}+\pi F \,,
\end{equation}
where we introduced the logarithm of the mirror S~matrix 
\begin{equation}
\label{eq:deltaphase}
\varphi_{12}(\bar{p}_1,\bar{p}_2)=-i\log \bar{S}(\bar{p}_1,\bar{p}_2)\,.
\end{equation}
In what follows, we will tacitly absorb the fermion number $F=\sum_j F_j$ in the phase shift $\varphi_{ab}$.
Note that $\beta$ is the volume of the theory and controls the spacing of its levels. Since we want to describe a thermal state in equilibrium we shall take the number of excitations~$M\sim\beta\gg1$. For each particle type~$a$ we introduce the density of occupation $\rho_{a}$, which counts the number of particles appearing in a given state. Introducing a (rapidity) parameter~$u$ which we shall specialise later, we have that $\rho_{a}(u)\sim \de N(p_{j,a})/(\beta\de u)$ is the number of occupied momenta in the state between $u$ and $u+\de u$ in units of $\beta$. The Bethe-Yang equations become
\begin{equation}
\bar{p}_{a}(u)+ \sum_{b=1}^{8+8} \int \de v\, \varphi_{ab}\big(\bar{p}_{a}(u),\bar{p}(v)\big)\,\rho_b(v)= 2\pi \frac{n_{a}}{\beta} \,.
\end{equation}
Our state is described by the levels $n_a$, and again it is convenient to consider their density $\rho'_a(u)\sim \de n_a/(\beta\,\de u)$. Introducing the kernel~$K_{ab}$
\begin{equation}
\label{eq:Kkernel}
K_{ab}(x,y)=\tfrac{1}{2\pi}\partial_{u}\varphi_{ab}(\bar{p}(u),\bar{p}(v))\,,
\end{equation}
 and the convolution~$*$
\begin{equation}
[K_{ab}*\rho_b](u)=\int \de v\, \sum_b K_{ab}(u,v)\,\rho_b(v)\,,\qquad
f_a*\rho_a=\int \de u\, \sum_a f_a(u)\,\rho_a(u)\,,
\end{equation}
 we have the familiar TBA constraint
\begin{equation}
\label{eq:statBA}
\frac{1}{2\pi}\partial_u\bar{p}_{a}(u) +  [K_{ab}*\rho_b](u)= \rho'_a(u) \,.
\end{equation}
Let us stress that $\rho'$ is the (total) density of levels, while $\rho$ is the density of occupation of such levels.
We can now evaluate the right-hand side of eq.~\eqref{eq:partition} at temperature~$1/R$ and in large volume, finding
\begin{equation}
\text{Tr}\left[ e^{-R\, \bar{\Hgen}}\right]=\int \mathcal{D}\rho_a\,\mathcal{D}\rho_a'\,
e^{-R\big(\bar{H} - \frac{1}{R} S\big)+i \pi F}\,.
\end{equation}
Using the densities, we have expressed the trace in terms of the mirror energy~$\bar{H}$ and of the entropy~$S$. Notice that we have introduced a chemical potential proportional to the fermion number~$F$ to account for the anti-periodicity of the fermions in the mirror model~\cite{Arutyunov:2007tc}. To find  the energy, it suffices to sum over the momenta of occupied states, which in terms of~$\rho_a$ gives
\begin{equation}
\bar{H} =\bar{H}_a*\rho_a\,.
\end{equation}
Similarly, the fermion number contribution is just
\begin{equation}
\label{eq:psi-a}
i\pi F=\psi_a *\rho_a\,,\qquad
\psi_a = \begin{cases}
0&a\in\text{bosons},\\
i\pi&a\in\text{fermions}.
\end{cases}
\end{equation}
The entropy is the logarithm of the number of states arising when $\{n_{a,j}\}_j$ energy levels are occupied by particles identified by $\{p_{a,j}\}_j$. This depends on whether a given level can be occupied by multiple particles (bosons) or by only one (fermions). The resulting combinatorial problem simplifies for large occupation numbers, and for bosons it gives
\begin{equation}
s^{\bo}[\rho,\rho']=\int \de u\,\Big[
(\rho'+\rho)\log(\rho'+\rho)- \rho\log\rho-\rho'\log\rho'
\Big],
\end{equation}
while for fermions
\begin{equation}
s^{\fe}[\rho,\rho']=\int \de u\,\Big[
\rho'\log\rho'- \rho\log\rho-(\rho'-\rho)\log(\rho'-\rho)
\Big].
\end{equation}
To find all distributions $\rho_a$ we just need to impose that the free energy~$\mathcal{F}$ is stationary,
\begin{equation}
\delta \mathcal{F}=\delta \Big(\bar{H}_a*\rho_a-\frac{1}{R}s[\rho_a,\rho'_a]+\frac{1}{R}\psi_a*\rho_a\Big) =0\,,
\end{equation}
subject to the constraint~\eqref{eq:statBA}. It is customary and convenient to write the resulting equations in terms of the \textit{pseudo-energies} $\epsilon_a(u)$ which are related to the filling fractions $\rho_a/\rho_a'$ as
\begin{equation}
\frac{\rho_a(u)}{\rho_a'(u)} = \frac{1}{e^{\epsilon_a(u)}-1}\quad\text{for bosons},\qquad
\frac{\rho_a(u)}{\rho_a'(u)} = \frac{1}{e^{\epsilon_a(u)}+1}\quad\text{for fermions}.
\end{equation}
The conditions for equilibrium are then
\begin{equation}
\epsilon_a(u)= \psi_a+R\, \bar{H}_a(u)-[\Lambda_b*K_{ba}](u)\,,\quad
\Lambda_b(v)=\begin{cases}
-\log(1-e^{-\epsilon_b(v)})& \text{bosons},\\
+\log(1+e^{-\epsilon_b(v)})& \text{fermions}.
\end{cases}
\end{equation}
and the free energy becomes
\begin{equation}
\mathcal{F}=-\frac{1}{2\pi R}\, (\partial_u \bar{p}_a)* \Lambda_a\,.
\end{equation}
At large $\beta$, the partition function~\eqref{eq:partition} is dominated by the contribution of $H_0$. On the other hand, in the same limit we have that $\log Z(\beta,R)= -\beta R\mathcal{F}(R)$. Hence we can read off the ground state energy \textit{of the original theory in finite volume},
\begin{equation}
H_0 = -\frac{1}{2\pi}\, (\partial_u\bar{p}_a)* \Lambda_a\,.
\end{equation}

\section{Wess-Zumino-Witten construction for \texorpdfstring{$\AdS{\mathbf{3}}$}{AdS3} strings}
\label{app:WZW}

The WZW model based on the $\mathcal N = 1$ Ka\v{c}-Moody algebras $\mathfrak{sl}(2)^{(1)}_k \oplus \mathfrak{su}(2)^{(1)}_k$ together with four free bosons and fermions provides a worldsheet description of closed strings propagating on $\AdSST$ with pure NS-NS flux~\cite{Giveon:1998ns,Maldacena:2000hw,Pakman:2003cu, Israel:2003ry,Raju:2007uj,Giribet:2007wp,Ferreira:2017pgt}. In the chiral sector, the modes of the $\mathfrak{sl}(2)^{(1)}_k$ currents are characterized by the following commutation relations 
\begin{equation}
\begin{aligned}
& [\text{L}^+_m, \text{L}^-_n] = - 2 \text{L}_{m+n}^3 + k m \delta_{m,-n} &  & [\text{L}^3_m, \text{L}^\pm_n] = \pm \text{L}^\pm_{m+n} & & [\text{L}^3_m, \text{L}^3_n] = -\frac{k}{2} m \delta_{m, -n}  \\
& [\text{L}_m^\pm, \psi_r^3] = \mp \psi_{r+m}^\pm & & [\text{L}_m^3, \psi_r^\pm] = \pm \psi_{r+m}^\pm & & [\text{L}_m^\pm, \psi_r^\mp] = \mp 2 \psi^3_{m+r} \\
& \{ \psi^+_r, \psi^-_s \} = k \delta_{r, -s} & & \{ \psi^3_r, \psi^3_s \} = -\frac{k}{2} \delta_{r, -s} \,,  
\end{aligned}
\label{eq:sl(2)-kac-moody-algebra}
\end{equation}
while for $\mathfrak{su}(2)^{(1)}_k$ modes we have 
\begin{equation}
\begin{aligned}
& [\text{J}^+_m, \text{J}^-_n] = 2 \text{J}_{m+n}^3 + k m \delta_{m,-n} &  & [\text{J}^3_m, \text{J}^\pm_n] = \pm \text{J}^\pm_{m+n} & & [\text{J}^3_m, \text{J}^3_n] = \frac{k}{2} m \delta_{m, -n}  \\
& [\text{J}_m^\pm, \theta_r^3] = \mp \theta_{r+m}^\pm & & [\text{J}_m^3, \theta_r^\pm] = \pm \theta_{r+m}^\pm & & [\text{J}_m^\pm, \theta_r^\mp] = \pm 2 \theta^3_{m+r} \\
& \{ \theta^+_r, \theta^-_s \} = k \delta_{r, -s} & & \{ \theta^3_r, \theta^3_s \} = \frac{k}{2} \delta_{r, -s} \,. 
\end{aligned}\label{eq:su(2)-kac-moody-algebra}
\end{equation}
The current zero-modes of the chiral algebra correspond to the global $\sl(2)_\L\oplus\su(2)_\L$ algebra of section~\ref{sec:symmetries}, \textit{i.e.}\ $L^I_0=\Lgen^I$ and $J^A_0=\Jgen^A$, while $\sl(2)_\R\oplus\su(2)_\R$ is given by the similar formulae in the anti-chiral sector.
As explained in detail in ref.~\cite{Ferreira:2017pgt}, in order to build the spectrum, it is convenient to define decoupled bosonic modes $L_n^a$ and $J_n^a$ that commute with the fermionic ones. As a result the level of the Ka\v{c}-Moody algebra is shifted by $+2$ and $-2$ respectively for $\mathfrak{sl}(2)$ and $\mathfrak{su}(2)$. Schematically,
\begin{equation}
\begin{aligned}
& \mathfrak{sl}(2)^{(1)}_k = \mathfrak{sl}(2)_{k+2} \oplus \text{free fermions} \,, \\
& \mathfrak{su}(2)^{(1)}_k = \mathfrak{su}(2)_{k-2} \oplus \text{free fermions} \,.
\end{aligned}
\end{equation}
In the following we will denote the uncharged free bosonic modes arising from the $\text{T}^4$ by $\alpha_n^{a \dot a}$ and their fermionic superpartners by $\gamma^{a \dot a}_n$. Worldsheet modes are charged under $\mathfrak{so}(4)_2 = \mathfrak{su}(2)_\bullet \oplus \mathfrak{su}(2)_\circ$. We list their charges in table~\ref{tab:circle-bullet-worldsheet}. What we have described so far is the chiral sector of the theory. We can similarly define the anti-chiral sector that will be denoted by tildes, \textit{e.g.}\ $\tilde L_n^{\dot{I}}$, $\tilde J_n^{\dot{A}}$, and so on. 

\begin{table}[t]
\centering
\renewcommand{\arraystretch}{1.2}
\begin{tabular}{|c|c|c|}
\hline
WZW mode & $\mathfrak{su}(2)_\bullet $ & $\mathfrak{su}(2)_\circ$  \\
\hline 
$L^\pm_n$ & $\boldsymbol 1$ & $\boldsymbol 1$ \\
$J^\pm_n$ & $\boldsymbol 1$ & $\boldsymbol 1$ \\
$\alpha^{a \dot a}_n$ & $\boldsymbol 2$ & $\boldsymbol 2$ \\
\hline
\end{tabular}
\hspace*{1cm}\begin{tabular}{|c|c|c|}
\hline
WZW mode & $\mathfrak{su}(2)_\bullet $ & $\mathfrak{su}(2)_\circ$  \\
\hline 
$\psi^\pm_n$ & $\boldsymbol 1$ & $\boldsymbol 1$ \\
$\theta^\pm_n$ & $\boldsymbol 1$ & $\boldsymbol 1$ \\
$\gamma^{a \dot a}_n$ & $\boldsymbol 2$ & $\boldsymbol 2$ \\
\hline
\end{tabular}
\renewcommand{\arraystretch}{1}
\caption{We list $\mathfrak{so}(4)_2 = \mathfrak{su}(2)_\bullet \oplus \mathfrak{su}(2)_\circ$ charges for worldsheet bosons and fermions respectively in the left and right table.}
\label{tab:circle-bullet-worldsheet}
\end{table}

To construct the spectrum of superstrings on $\AdSST$ we need to consider fermions on the worldsheet with periodic boundary conditions in the Ramond (R)  sector, and anti-periodic ones in the Neveu-Schwarz (NS) sector. As a consequence, the fermions are integer-moded in the R sector and half-integer-moded in the NS sector. Target-space supersymmetry will follow from the GSO projection.

The $\sl(2)$ WZW model presents a further complication: the physical spectrum is constructed from representations arising from different \textit{spectrally flowed} sectors \cite{Maldacena:2000hw}. We will briefly review such construction below, starting from the simplest sector where no spectral flow is needed.

\subsection{The ``spectrally unflowed'' representations}
The simplest part of the spectrum, which turns out to describe states whose energy is small with respect to the WZW level, is the so-called unflowed sector. Here to construct a physical state we start from a highest-weight Ka\v{c}-Moody module. This is identified by a lowest-weight state of $\mathfrak{sl}(2)$ and highest-weight state of $\su(2)$, which we label~$\ket{\ell_0,j_0}$. Acting with $\sl(2)_k$ Ka\v{c}-Moody algebra we have
\begin{equation}
L_0^- \ket{\ell_0,j_0} = 0 \,, \qquad L_0^3 \ket{\ell_0,j_0} = \ell_0 \ket{\ell_0,j_0} \,, \qquad  L_n^a \ket{\ell_0,j_0} = 0 \quad \text{for} \ n> 0 \,. 
\label{eq:unflowed-rep-sl}
\end{equation} 
while for $\su(2)_k$
\begin{equation}
\label{eq:unflowed-rep-su}
J_0^+ \ket{\ell_0,j_0} = 0 \,, \qquad J_0^3 \ket{\ell_0,j_0} = j_0 \ket{j_0} \,, \qquad J_n^a \ket{\ell_0,j_0} = 0 \quad \text{for} \ n> 0 \,.
\end{equation} 
We can construct descendants of the global symmetry algebra $\sl(2)_\L\oplus\su(2)_\L$ by acting with raising and lowering operators. For $\sl(2)_\L$ we have infinite-dimensional representations
\begin{equation}
L_0^3 (L_0^+)^p \ket{\ell_0} = (\ell_0 + p) \ket{\ell_0} \,, \qquad p \geq 0 \,,
\end{equation}
whereas of course $\su(2)_\L$ representations are finite-dimensional with dimension $2 j_0+1$:
\begin{equation}
J_0^3 (J_0^-)^p \ket{j_0} = (j_0 - p) \ket{j_0} \,, \qquad 0 \leq p \leq 2 j_0 \,. 
\end{equation} 
which results in null-state conditions on the $\su(2)$ part of the spectrum.
Generic states in the Ka\v{c}-Moody module can be created by acting with negative-moded modes of the currents, $L^\pm_{-n}$ and $J^\pm_{-n}$, $n\geq1$.

Unitarity together with some observations by Maldacena and Ooguri~\cite{Maldacena:2000hw} give constraints on the possible values of $\ell_0$ and $j_0$ for such representations:
\begin{equation}
\dfrac{1}{2} < \ell_0 < \dfrac{k+1}{2} \,, \qquad 0 \leq j_0 \leq \dfrac{k-2}{2} \,. 
\label{eq:spin-bounds}
\end{equation}
It is convenient to introduce an ``effective'' number of excitations,
\begin{equation}
\label{eq:Neff}
N_{\rm{eff}} = \begin{cases} \hat{N} - \tfrac{1}{2} & \rm{NS \ sector} \\ \hat{N} & \rm{R \ sector}  \end{cases}, 
\end{equation}
where $\hat{N}$ is the eigenvalue of the total number operator.
Then the \textit{mass-shell condition} which physical states should obey, in the chiral sector reads
\begin{equation}
-\dfrac{\ell_0(\ell_0 - 1)}{k} + \dfrac{j_0(j_0 + 1)}{k} + \Neff = 0. 
\label{eq:unflowed-mass-shell}
\end{equation}
Notice that eqs.~\eqref{eq:spin-bounds} together with \eqref{eq:unflowed-mass-shell} constrain the values of $\ell_0$ and of $\Neff$ to be bounded. Indeed only part of the spectrum is described by the ``unflowed'' representations; we shall see below how to describe more general representations.

To get a physical state, we still need to tensor together the chiral and anti-chiral representations and impose the GSO projection.
Firstly, we should take the same $\mathfrak{sl}(2)$ representation in both the chiral and anti-chiral sector, so that $\ell_0 = \tilde \ell_0$~\cite{Maldacena:2000hw}. Similarly, for the $\mathfrak{su}(2)$ representation we have $j_0 = \tilde \jmath_0$. Level-matching requires 
\begin{equation}
\NN = \NNR \,. 
\label{eq:unflowed-level-matching}
\end{equation}
The GSO projection in the NS sector simply amounts to requiring that $\NN$ is integer, while in the R sector it requires an even number of fermions. However, in practice in the latter case things are somewhat more involved due to the presence of zero modes as we shall see just below, see also appendix~\ref{app:WZWBPS} for some examples.

To make contact with the target-space charges of section~\ref{sec:symmetries}, let us denote by $\ell$ the eigenvalue of $L^3_0=\Lgen^3$ on a generic state, and by $j$ the eigenvalue of $J^3_0=\Jgen^3$ (and similarly in the anti-chiral sector). Then the total light-cone energe~$\Htot$ is given by the sum of the chiral and anti-chiral contributions: 
\begin{equation}
\Htot = \ell-j + \tilde{\ell}- \tilde{\jmath} \,.
\end{equation}
In terms of the quantum numbers of $|\ell_0,j_0\rangle$, we write that
\begin{align}
& \ell = \ell_0 + \delta \ell \,,   \qquad \quad  j = j_0  - \delta j \,,  & \rm{NS \ sector} \label{eq:unflowed-spins-NS} \\ 
& \ell = \ell_0 + \delta \ell + s_{\ell} \,,  \quad \; j = j_0  - \delta j - s_j \,, & \rm{R \ sector} 
\label{eq:unflowed-spins-R}
\end{align}
We introduced the notation $\delta \ell$ to denote the $\sl(2)$ charge of the state with respect to the ground state $|\ell_0,j_0\rangle$. This is equal to the number of $L^+_{-n}$ modes minus the number of $L^-_{-n}$ modes used in the construction of the state. Note that in our convention each $\mathfrak{su}(2)$ mode~$J^\pm_{-n}$ contributes to $\delta j$ with $\mp 1$. In the R sector, we have four additional labels $s_{\ell}, s_j, s', s'' = \pm \tfrac{1}{2}$, which identify the different choices of fermionic zero-modes of the R-sector vacuum, as described in \textit{e.g.}~\cite{Ferreira:2017pgt}.%
\footnote{Notice that in terms of these numbers the GSO projection imposes $s_{\ell} - s_j + s' + s'' + F \in 2 \mathbb{Z}$ in the R sector, where $F$ denotes the contribution to worldsheet fermion number coming from all but the zero-mode fermions.}
With these identifications in mind, we can write down the light-cone energy~$H$ of a state satisfying the mass-shell condition~\eqref{eq:unflowed-mass-shell}. To this end, we should see the latter as a (quadratic) equation on the allowed $\ell_0$, given $j_0$ and the total excitation level~$\Neff$. Then
\begin{equation}
\label{eq:unflowed-mass-shell-solution}
\Htot=\sqrt{(2j_0+1)^2+4k\Neff}-(2j_0+1)+\delta +\tilde{\delta}\,,
\end{equation}
with
\begin{equation}
\label{eq:deltadef}
\delta=\delta \ell+\delta j+s_{\ell}+s_j+1\,,\qquad
\tilde{\delta}=\delta\tilde{\ell}+\tilde{s}_j+\delta\tilde{\jmath} +\tilde{s}_j+1.
\end{equation}
 Notice that in solving the quadratic equation~\eqref{eq:unflowed-mass-shell} we have selected the positive root, as required by eq.~\eqref{eq:spin-bounds}. Below we shall illustrate these rules by constructing some low-lying physical states in the $\ket{\ell_0,j_0}$ module.

\renewcommand{\arraystretch}{1.5}
\begin{table}
\centering
\begin{tabular}{|c|c|c|c|c|c|}
\hline
WZW state & $\ell$ & $j$ & $\ell - j$ & $J_{\bullet}$ & $J_{\circ}$ \\
\hline 
\hline
$\psi^-_{-1/2} \ket{\ell_0, \ell_0 - 1}$ & $\ell_0 -1$ & $\ell_0 -1$ & 0 & 0 & 0 \\
\hline
$\theta^+_{-1/2} \ket{\ell_0, \ell_0 - 1}$ & $\ell_0$ & $\ell_0 $ & 0  & 0 & 0 \\
\hline
$\gamma^{a \dot a}_{-1/2} \ket{\ell_0, \ell_0 - 1}$ & $\ell_0$ & $\ell_0 -1$ & 1 & $\pm \frac{1}{2}$ & $\pm \frac{1}{2}$ \\
\hline
$\theta^-_{-1/2} \ket{\ell_0, \ell_0 - 1}$ & $\ell_0$ & $\ell_0 -2 $ & 2 & 0 & 0 \\
\hline
$\psi^+_{-1/2} \ket{\ell_0, \ell_0 - 1}$ & $\ell_0 + 1$ & $\ell_0 -1$ & 2 & 0 & 0 \\
\hline 
\end{tabular}
\caption{Level $\NN =0$ states in the NS sector. The first two states satisfy the BPS condition $\ell=j$; the remaining six states are not BPS.}
\label{tab:level-zero-NS}
\end{table}
\renewcommand{\arraystretch}{1}

\subsection{Some low-lying states in the unflowed sector}
\label{app:WZWBPS}
In the unflowed sector BPS states occur at level $\NN = 0$. The mass-shell condition \eqref{eq:unflowed-mass-shell} implies $j_0 = \ell_0-1$. In the NS chiral sector we find two BPS states: 
\begin{equation}
\begin{aligned}
& \psi_{-\frac{1}{2}}^- \ket{\ell_0, \ell_0 -1} \,, \qquad \delta \ell = -1 \,, \quad \delta j = 0 \,, \quad \ell = j = \ell_0 -1  \\ 
& \theta_{-\frac{1}{2}}^+ \ket{\ell_0, \ell_0 -1} \,, \qquad \delta \ell = 0 \,, \quad \delta j = -1 \,, \quad \ell = j = \ell_0  
\end{aligned}
\end{equation}
for $\ell_0 = 1, \dots, \tfrac{k}{2}$. Notice that they obey the GSO projection. BPS states can be built in the anti-chiral NS sector in a similar way. We find other two BPS states in the R sector: 
\begin{equation}
\begin{aligned}
& \ket{\ell_0, \ell_0 -1}_{(\, - \, - \, + \, -)} \,, \qquad \delta \ell = \delta j = 0 \,, \quad \ell = j = \ell_0 -\frac{1}{2}  \,, \\ 
& \ket{\ell_0, \ell_0 -1}_{(\, - \, - \, - \, +)} \,, \qquad \delta \ell = \delta j = 0 \,, \quad \ell = j = \ell_0 -\frac{1}{2} \,,
\end{aligned}
\label{eq:R-sector-BPS-states}
\end{equation}
where the subscript $(\, \pm, \, \pm, \, \pm, \, \pm \, )$ denotes the sixteen combinations of $(s_{\ell}, s_j, s', s'') = (\, \pm \frac{1}{2}, \, \pm \frac{1}{2}, \, \pm \frac{1}{2}, \, \pm \frac{1}{2} \, )$. Notice that the BPS states in eq. \eqref{eq:R-sector-BPS-states} have $s_{\ell} - s_j + s' + s'' = 0$, hence obeying the GSO projection. Combining R and NS, chiral and anti-chiral sectors we find a total of sixteen BPS states for $\ell_0 = j_0-1 = 1, \dots, \tfrac{k}{2}$. For each of these values they give rise to the Hodge diamond of $\text{T}^4$: representing a state by its $(\ell=j;\tilde{\ell}=\tilde{\jmath})$ charges we have, for a given $j_0$,
\begin{equation}
\label{eq:HodgeDiamond}
\begin{matrix} 
 &  & (j_0,j_0) & & \\
 & (j_0+\tfrac{1}{2},j_0)^{\dot{a}} &  & (j_0,j_0+\tfrac{1}{2})^{\dot{a}} & \\
 (j_0+1,j_0) & & (j_0+\tfrac{1}{2},j_0+\tfrac{1}{2})^{\dot{a}\dot{b}} & & (j_0,j_0+1)  \\
 & (j_0+1,j_0+\tfrac{1}{2})^{\dot{a}} &  & (j_0+\tfrac{1}{2},j_0+1)^{\dot{a}}  & \\ 
 &  & (j_0+1,j_0+1) & & 
\end{matrix}
\end{equation} 
The dotted indices $\dot{a},\dot{b}$ take values $1$ or $2$ and in fact give the fundamental representation of~$\su(2)_\circ$ in the decomposition of section~\ref{sec:symmetries}.

At level $\NN = 0$ there exist further (non-BPS) states. We shall list those separately for the NS and R sector. Notice that in both sectors, $\NN = 0$ together with the mass-shell condition \eqref{eq:unflowed-mass-shell} and eq.~\eqref{eq:spin-bounds} implies $\ell_0 = j_0 +1$.

\paragraph{NS sector.}
Since $\NN =0$, we have exactly one fermion acting on the ground state $\ket{\ell_0, \ell_0 -1}$. We find eight physical states, that are listed in Table \ref{tab:level-zero-NS} together with their charges.  The construction of states at higher level works similarly: for each fixed $\NN \geq 0$ one solves the mass-shell condition \eqref{eq:unflowed-mass-shell} for $\ell_0$ as a function of $j_0$, paying attention to respect the bounds in eq. \eqref{eq:spin-bounds}. Negative-moded bosonic and fermionic operators are then applied to the ground state $\ket{\ell_0, j_0}$ to reach the chosen value of $\NN$.

\paragraph{R sector.}

The R sector is characterized by the presence of fermionic zero-modes. As described in \cite{Ferreira:2017pgt} their action can be accounted for introducing the four quantum numbers $s_{\ell}, s_j, s', s'' = \pm \tfrac{1}{2}$. At level $\NN = 0$, the GSO projection imposes $s_{\ell} - s_j + s' + s'' \in  2 \mathbb{Z}$.%
\footnote{%
The sign in front of $s_j$ is different because we identified the $\su(2)$ charge oppositely with respect to the $\sl(2)$ one. 
}
 We find eight physical states of the form $\ket{\ell_0, \ell_0 - 1}_{(s_{\ell}, s_j, s', s'')}$ that we list in Table \ref{tab:level-zero-R}. Notice that different choices of $(s_{\ell}, s_j, s', s'')$ carry different $\mathfrak{so}(4)_2 = \mathfrak{su}(2)_\bullet \oplus \mathfrak{su}(2)_\circ$ charges. In particular we find 
\begin{equation}
\label{eq:sjrelation}
J_\bullet = \dfrac{s' + s''}{2} \,, \qquad J_\circ = \dfrac{s'' - s'}{2} \,. 
\end{equation}
The construction of physical states at higher $\NN$ level proceeds as in the NS sector. The only difference is in the GSO projection that imposes to choose the eight allowed combinations of $(s_{\ell}, s_j, s', s'')$ differently for even and odd worldsheet fermion number $F$. 

\renewcommand{\arraystretch}{1.5}
\begin{table}
\centering
\begin{tabular}{|c|c|c|c|c|c|}
\hline
$(s_{\ell}, s_j, s',s'')$ & $\ell$ & $j$ & $\ell - j$ & $J_{\bullet}$ & $J_{\circ}$\\
\hline
\hline 
$(\, - \, - \, - \, + \, )$  & $\ell_0 - \frac{1}{2}$ & $\ell_0 - \frac{1}{2}$ & 0 & 0 & $+\frac{1}{2}$ \\
\hline
$(\, - \, - \, + \, - \,)$  & $\ell_0 - \frac{1}{2}$ & $\ell_0 - \frac{1}{2}$ & 0 & 0 & $-\frac{1}{2}$\\
\hline 
$(\, - \, + \, - \, - \,)$  & $\ell_0 - \frac{1}{2}$ & $\ell_0 - \frac{3}{2}$ & 1 & $-\frac{1}{2}$ & 0 \\
\hline 
$(\, - \, + \, + \, +\,)$ & $\ell_0 - \frac{1}{2}$ & $\ell_0 - \frac{3}{2}$ & 1 & $+\frac{1}{2}$ & 0 \\
\hline 
$(\, + \,- \, - \, - \,)$ & $\ell_0 + \frac{1}{2}$ & $\ell_0 - \frac{1}{2}$ & 1 & $-\frac{1}{2}$ & 0\\
\hline 
$(\, + \, - \, + \, + \,)$  & $\ell_0 + \frac{1}{2}$ & $\ell_0 - \frac{1}{2}$ & 1  & $+\frac{1}{2}$ & 0 \\
\hline
$(\, + \, + \, - \, + \,)$  & $\ell_0 + \frac{1}{2}$ & $\ell_0 - \frac{3}{2}$ & 2 & 0 & $+\frac{1}{2}$\\
\hline 
$(\, + \, + \, + \, - \, )$  & $\ell_0 + \frac{1}{2}$ & $\ell_0 - \frac{3}{2}$ & 2 & 0 & $-\frac{1}{2}$  \\
\hline
\end{tabular}
\caption{Level $\NN =0$ states in the R sector. In the first column we adopt the shorthand notation $(\, \pm \,, \pm \,, \pm \,, \pm \,)$ for the state $\ket{\ell_0, \ell_0-1}_{(s_{\ell}, s_j, s', s'')}$ with $(s_{\ell}, s_j, s', s'') = \left( \, \pm \tfrac{1}{2} \,, \, \pm \tfrac{1}{2} \,, \, \pm \tfrac{1}{2} \,, \, \pm \tfrac{1}{2} \,  \right)$.  }
\label{tab:level-zero-R}
\end{table}
\renewcommand{\arraystretch}{1}

\subsection{Spectral flow}
\label{sec:flowed-sector}

The Maldacena-Ooguri bound \eqref{eq:spin-bounds} constrains the maximal spacetime energy for states arising from the unflowed sector. This gives much fewer states than what expected. To resolve this issue Maldacena and Ooguri argued in ref.~\cite{Maldacena:2000hw} that the spectrum contains also additional $\mathfrak{sl}(2)_k$ representations; the so-called spectrally flowed representations. Spectrally flowed representations can also be defined for $\mathfrak{su}(2)_k$. However their nature is quite different: while for $\mathfrak{sl}(2)_k$ spectral flow produces new inequivalent representations, in the case of $\mathfrak{su}(2)_k$  this is not the case. Rather, spectral flow in  $\mathfrak{su}(2)_k$ can be a convenient way to relabel states within the same module. Below we briefly review spectrally flowed $\mathfrak{sl}(2)_k$ and $\mathfrak{su}(2)_k$ representations. 

\paragraph{Spectral flow in $\mathfrak{sl}(2)$.}

Spectral flow is an outer automorphism $\pi_w$ of the Ka\v{c}-Moody algebra in \eqref{eq:sl(2)-kac-moody-algebra}. It is defined for any $w \in \mathbb{Z}$ by the following relations
\begin{equation}
\begin{aligned}
& \pi_w(\text{L}_n^\pm) := \hat{\text{L}}_n^\pm = \text{L}^\pm_{n \pm w}\ ,\\
& \pi_w(\text{L}_n^3):= \hat{\text{L}}_n^3 = \text{L}^3_n - \frac{k}{2}w \, \delta_{n,0}\ ,   \\
& \pi_w(\mathcal L_n^{\mathfrak{sl}}) := \hat{\mathcal L}_n^{\mathfrak{sl}} = \mathcal L_n^{\mathfrak{sl}} + w \, \text{L}_n^3 - \dfrac{k}{4}w^2 \delta_{n,0}\ , \\
& \pi_w(\psi_r^3) := \hat \psi_r^3 = \psi_r^3\ , \\
& \pi_w(\psi_r^\pm) := \hat \psi_r^\pm = \psi^\pm_{r \pm w} \ .  
\end{aligned}
\label{eq:spectrally-flowed-sl2-generators}
\end{equation}
Indeed, one can verify that the commutation relations in eq.~\eqref{eq:sl(2)-kac-moody-algebra} are preserved. 
Spectrally flowed representations are the image under the spectral flow automorphism \eqref{eq:spectrally-flowed-sl2-generators} of unflowed representations. Applying the spectral flow automorphism $\pi_w$ on the two sides of \eqref{eq:unflowed-rep-sl} we find new $\mathfrak{sl}(2)_k$ representations, defined by the following relations:
\begin{equation}
\begin{gathered}
L^+_{n +  w} \ket{\ell_0;  w} = 0 \,, \qquad L^-_{n - w - 1} \ket{\ell_0;  w} = 0 \,, \qquad L_n^3 \ket{\ell_0;  w} = 0 \,, \qquad n = 1, 2, \dots\,, \\
 L_0^3 \ket{\ell_0;  w} = \left( \ell_0 + \frac{kw}{2}  \right) \ket{\ell_0;  w}\,.
\end{gathered}
\label{eq:flowed-rep-sl2}
\end{equation}
Since we want to decompose the spectrally flowed representations in terms of lowest weight representations of the global $\mathfrak{sl}(2)$, we need
\begin{equation}
L_0^- \ket{\ell_0;  w} = \hat{L}_{ w}^- \ket{\ell_0; w}  = 0 \,. 
\end{equation}
We will therefore assume from now on $ w > 0$. Since $\mathfrak{sl}(2, \mathbb{R})$ is a non-compact algebra, the representations defined in \eqref{eq:flowed-rep-sl2} are inequivalent representations. In particular remark that for spectrally flowed representations the eigenvalue of the Virasoro generator $\mathcal{L}_0^{\sl}$, the conformal dimension, is unbounded from below. For example, 
\begin{equation}
\begin{gathered}
(L_ w^+)^p \ket{\ell_0;  w} \neq 0 \quad  \forall p \geq 0\,, \\
\mathcal{L}_0^{\mathfrak{sl}}\, (L_ w^+)^p \ket{\ell_0;  w} = \Big[-  w \, p - \frac{k}{4}w^2-\frac{\ell_0(\ell_0-1)}{k}\Big] (L_ w^+)^p \ket{\ell_0;  w}\,.
\end{gathered}
\end{equation}
The (left) target space energy~$\ell$ (\textit{i.e.}, the eigenvalue of ~$\Lgen^3=L^3_0$) can be computed using
\begin{equation}
\begin{aligned}
& \ell = \ell_0 + \delta \ell + \dfrac{kw}{2} \,, \qquad & \text{NS \ sector,} \\
& \ell = \ell_0 + \delta \ell + s_{\ell} + \dfrac{kw}{2} \,, \qquad & \text{R \ sector.} \\
\end{aligned}
\end{equation}

\paragraph{Spectral flow in $\mathfrak{su}(2)$.}

Even though this will not produce new representations, in the following we will find it convenient to spectral flow also the $\mathfrak{su}(2)$ algebra. Similarly to the $\mathfrak{sl}(2)$ case a spectral flow automorphism can be defined for the algebra in eq.~\eqref{eq:su(2)-kac-moody-algebra}:
\begin{equation}
\begin{aligned}
& \pi_w(\text{J}_n^\pm) := \hat{\text{J}}_n^\pm = \text{J}^\pm_{n \mp w}\ ,\\
& \pi_w(\text{J}_n^3):= \hat{\text{J}}_n^3 = \text{J}^3_n - \frac{k}{2}w \, \delta_{n,0}\ ,   \\
& \pi_w(\mathcal L_n^{\mathfrak{su}}) := \hat{\mathcal L}_n^{\mathfrak{su}} = \mathcal L_n^{\mathfrak{su}} - w \, \text{J}_n^3 + \dfrac{k}{4}w^2 \delta_{n,0}\ , \\
& \pi_w(\theta_r^3) := \hat \theta_r^3 = \theta_r^3\ , \\
& \pi_w(\theta_r^\pm) := \hat \theta_r^\pm = \theta^\pm_{r \mp w} \ .  
\end{aligned}
\label{eq:spectrally-flowed-su2-generators}
\end{equation}
Also in this case one can define spectrally flowed representations according to 
\begin{equation}
\begin{aligned}
& J^-_{n + \omega} \ket{j_0; \omega} = 0 \,, \quad J^+_{n - w - 1} \ket{j_0; \omega} = 0 \,, \quad J_n^3 \ket{\ell_0; \omega} = 0 \,, \qquad n = 1, 2, \dots \\
& J_0^3 \ket{j_0; \omega} = \left( j_0 + \frac{kw}{2}  \right) \ket{j_0; \omega}
\end{aligned}
\label{eq:flowed-rep-su2}
\end{equation}
However, it can be shown (see \textit{e.g.}~\cite{Giribet:2007wp}) that this does not produce new representations: for $w$ even (resp.\ odd), a spin $j_0$ representation is mapped to a spin $j_0$ (resp.\ $\tfrac{k}{2} - j_0$) representation. In particular we have, in the NS sector \cite{Pakman:2003cu,Giribet:2007wp}
\begin{equation}
\ket{j_0;w} = \theta^+_{-w+\frac{1}{2}} \dots \theta^+_{-\frac{1}{2}} (J^+_{-w})^{2j_0} (J^+_{-w +1})^{k'-2j_0} \dots (J^+_{-2})^{2j_0} (J^+_{-1})^{k' - 2j_0} \ket{j_0} \,,
\end{equation}
for $w$ even and 
\begin{equation}
\ket{j_0;w} = \theta^+_{-w+\frac{1}{2}} \dots \theta^+_{-\frac{1}{2}} (J^+_{-w})^{2j_0} (J^+_{-w +1})^{k'-2j_0} \dots (J^+_{-2})^{k'-2j_0} (J^+_{-1})^{2j_0} \ket{\tfrac{k}{2}-j_0} \,, 
\label{eq:flowed-su2-states}
\end{equation}
for $w$ odd, where $k' = k-2$ is the level of the decoupled bosonic $\mathfrak{su}(2)$ algebra. Similar expressions hold in the R sector. The expression for $j$ and $\tilde \jmath$ in eqs.~\eqref{eq:unflowed-spins-NS} and \eqref{eq:unflowed-spins-R} should now be replaced by 
\begin{equation}
\begin{aligned}
& j = j_0 - \delta j + \dfrac{kw}{2} \,, \qquad & \text{NS \ sector} \,, \\
& j = j_0 - \delta j - s_j + \dfrac{kw}{2} \,, \qquad & \text{R \ sector} \,. 
\end{aligned}
\end{equation}

\paragraph{Spectral flow and target-space charges.}

Let us see how the physical-state condition changes when performing spectral flow by the same $w > 0$ in both $\mathfrak{sl}(2)_k$ and $\mathfrak{su}(2)_k$. This is a particularly convenient choice as BPS states in the unflowed representation are mapped to  BPS states in flowed representations. Indeed,  
\begin{equation}
\hat L^3_0 - \hat J^3_0 = L^3_0 - J^3_0 \,. 
\end{equation}
Due to the choice of equal spectral flow parameter for both $\mathfrak{sl}(2)_k$ and $\mathfrak{su}(2)_k$, the mass-shell condition in the spectrally flowed sector takes the simple form \cite{Ferreira:2017pgt}
\begin{equation}
-\dfrac{\ell_0(\ell_0-1)}{k} - w(\ell_0 + \delta \ell) + \dfrac{j_0(j_0+1)}{k} + w(j_0 - \delta j) + \NN = 0
\label{eq:flowed-mass-shell-NS}
\end{equation}
in the NS sector and 
\begin{equation}
-\dfrac{\ell_0(\ell_0-1)}{k} - w(\ell_0 + \delta \ell + s_{\ell}) + \dfrac{j_0(j_0+1)}{k} + w(j_0 - \delta j - s_j) + \NN = 0 \,. 
\label{eq:flowed-mass-shell-R}
\end{equation}
in the R sector. Similar expressions hold in the anti-chiral sector. The level-matching condition can be deduced from the mass-shell condition imposing $\ell_0 = \tilde \ell_0$ and $j_0 = \tilde \jmath_0$. We find
\begin{equation}
\NN - \NNR = w (\delta - \tilde \delta) \,,
\label{eq:flowed-level-matching}
\end{equation} 
where $\delta, \tilde{\delta}$ are given by~eq.~\eqref{eq:deltadef} and again we understand that~$s_{\ell}$ and $s_j$ are zero in the NS sector.
Finally, since each unit of spectral flow contributes to the fermion number by two (one from $\mathfrak{sl}(2)$ and one from $\mathfrak{su}(2)$) the GSO projection takes the same form it took in the unflowed sector. In conclusion we can write the light-cone energy~$H$ as
\begin{equation}
\label{eq:flowed-mass-shell-solution}
\Htot=\sqrt{(2j_0+k w+1)^2 + 4 k (\Neff- w \delta)}-(2j_0+k w+1)+\delta +\tilde{\delta},
\end{equation}
which takes the same form as eq.~\eqref{eq:unflowed-mass-shell-solution} with shifted variables.

\paragraph{Asymmetric spectral flow.}

As we have discussed, the spectral flow automorphism does not generate new representations of~$\su(2)_k$. Moreover a spin~$j_0$ representation is mapped onto itself provided that the spectral flow parameter~$w$ is~even. It is therefore possible to perform the spectral flow differently in the chiral and anti-chiral $\su(2)_k$ representations,%
\footnote{%
As discussed above, this is not possible for $\sl(2)_k$, where we should use the same spectral flow parameter~$w$ in the chiral and anti-chiral sectors.}
 provided that the difference of the two spectral-flow parameters $w_L-w_R$ is even. In view of the discussions above, it is conceptually straightforward (though somewhat cumbersome) to work out the solution to the mass shell condition for such an asymmetric spectral flow. Let us do so in the NS sector---the R sector works analogously.
The mass-shell condition in the chiral sector is 
\begin{equation}
- \dfrac{\ell_0(\ell_0 - 1)}{k} - w(\ell_0 + \delta \ell) - \dfrac{k w^2}{4} + \dfrac{j_0(j_0 + 1)}{k} + w_L(j_0 - \delta j) + \dfrac{k w_L^2}{4} + \NN = 0\,,
\label{eq:l0-asymmetric}
\end{equation} 
while in the anti-chiral sector
\begin{equation}
- \dfrac{\ell_0(\ell_0 - 1)}{k} - w(\ell_0 + \delta \tilde \ell) - \dfrac{k w^2}{4} + \dfrac{j_0(j_0 + 1)}{k} + w_R(j_0 - \delta \tilde \jmath) + \dfrac{k w_R^2}{4} + \NNR = 0\,.
\end{equation} 
The level-matching condition reads 
\begin{equation}
\NN -\NNR =  w (\delta \ell - \delta \tilde \ell) - w_R \delta \tilde \jmath + w_L \delta j + \dfrac{w_R-w_L}{2} \left( 2 j_0  + k \dfrac{w_R+w_L}{2} \right).
\end{equation}
Solving eq. \eqref{eq:l0-asymmetric} for $\ell_0$ we find
\begin{equation}
\ell_0 = \dfrac{1}{2} - \dfrac{kw}{2} + \dfrac{1}{2} \sqrt{1 + 4 j_0 + 4 j_0^2 +  4 k \NN - 2 k w - 4 k w \delta \ell - 4 k w_L (\delta j - j_0) + k^2 w_L^2}\,.
\end{equation}
Let us now specialise our analysis to the case
\begin{equation}
w = \dfrac{w_L + w_R}{2}\,.
\end{equation}
This means that the ``average'' of the $\su(2)_k$ spectral flow coincides with the $\sl(2)_k$ spectral flow. This is the closest analogue to the ``supersymmetric'' spectral flow considered above. Notice that indeed $\tfrac{1}{2}(w_L+w_R)\in\mathbb{N}$, as we are requiring the difference~$w_L-w_R$ to be even.
Using the level-matching condition we can rewrite
\begin{equation}
\ell_0 = \dfrac{1}{2} - \dfrac{kw}{2} + \dfrac{1}{2} \sqrt{ \Big(2j_0 + 1 + \frac{w_L + w_R}{2}k \Big)^2 + 2 k (\mathcal{N} +   \tilde{\mathcal{N}}) + k^2 \dfrac{(w_L - w_R)^2}{4} }\,,
\end{equation}
where we have introduced the short-hand notations
\begin{equation}
\begin{aligned}
\mathcal{N} = \NN - w \big(\delta\ell +\frac{1}{2}\big)  - w_L \big(\delta j +\frac{1}{2}\big)\,, \\
\tilde{\mathcal{N}} = \NNR - w \big(\delta\tilde{\ell} +\frac{1}{2}\big)  - w_L \big(\delta \tilde{\jmath} +\frac{1}{2}\big)\,.
\end{aligned}
\end{equation}
Notice that the quantities $\Nsc$ and $\tilde{\Nsc}$ are integer owing to the GSO projection~\cite{Ferreira:2017pgt}.
Putting together the chiral and anti-chiral contributions, we can write the light-cone energy as
\begin{equation}
\begin{aligned}
\Htot&=&\sqrt{ \Big(2j_0 + 1 + \frac{w_L + w_R}{2}k \Big)^2 + 2 k (\mathcal{N} +   \tilde{\mathcal{N}}) + k^2 \dfrac{(w_L - w_R)^2}{4} }
\qquad\qquad\qquad\\
&&-
\Big(2j_0 + 1 + \frac{w_L + w_R}{2}k \Big)+\delta+\tilde{\delta}\,,
\end{aligned}
\end{equation}
where $\delta$ and $\tilde{\delta}$ are again given by eq.~\eqref{eq:deltadef}.
Exploiting once more the level-matching condition we also find 
\begin{equation}
\mathcal{N} - \tilde{\mathcal{N}} = \dfrac{w_R-w_L}{2} \left( 2 j_0  + k \dfrac{w_R+w_L}{2} + 1 \right)\,.
\end{equation}

\bibliographystyle{JHEP}
\bibliography{refs}

\end{document}